\input epsf.tex
\documentstyle[natbib209,2up,multicol,float]{arPrepn}
\twouparticle

\newcommand{\sze}{SZE}

\newcommand{\Ho}{\mbox{$H_0$}}         %Hubble constant
\newcommand{\neo}{\mbox{$n_{e  0}$}}   % central density
\newcommand{\dTo}{\mbox{$\Delta T_0$}} % central decrement/increment
\newcommand{\Sxo}{\mbox{$S_{x  0}$}}   % central surface brightness
\newcommand{\Sx}{\mbox{$S_{x}$}}       % surface brightness
\newcommand{\Lamo}{\mbox{$\Lambda_{e \mbox{\tiny H} 0}$}}
\newcommand{\LameH}{\mbox{$\Lambda_{e \mbox{\tiny H}}$}}

\newcommand{\Tcmb}{\mbox{$T_{CMB}$}}     % T_CMB
\newcommand{\Da}{\mbox{$D_{\!\mbox{\tiny A}}$}}       % angular diam dist
\newcommand{\kms}{\mbox{km s$^{-1}$}}
\newcommand{\ksM}{\mbox{km s$^{-1}$ Mpc$^{-1}$}}

\newcommand{\lsim}{\lesssim}

\newcommand{\Om}{\mbox{$\Omega_M$}}
\newcommand{\Ol}{\mbox{$\Omega_\Lambda$}}

\newcommand{\Ob}{\mbox{$\Omega_B$}}
\newcommand{\OmM}{\Om}
\newcommand{\OmL}{\Ol}

\newcommand{\OmB}{\Ob}

\newcommand{\fb}{\mbox{$f_B$}}	      % baryon fraction

\newcommand{\kb}{\mbox{$k_{\mbox{\tiny B}}$}}     % Boltzmann's constant
\newcommand{\kB}{\kb}

\floatstyle{plain}
\newfloat{plate}{thp}{pla}
\floatname{plate}{Plate}

\renewcommand{\refname}{Literature Cited}

\newcommand{\aj}{AJ}%          % Astronomical Journal 
\newcommand{\araa}{ARA\&A}%    % Annual Review of Astron and Astrophys 
\newcommand{\apj}{ApJ}%        % Astrophysical Journal 
\newcommand{\apjl}{ApJ}%{ApJL?}% Astrophysical Journal, Letters 
\newcommand{\apjs}{ApJS}% 
\newcommand{\apss}{Ap\&SS}%    % Astrophysics and Space Science 
\newcommand{\aap}{A\&A}%       % Astronomy and Astrophysics 
\newcommand{\mnras}{MNRAS}%   % Monthly Notices of the RAS 
\newcommand{\prd}{Phys.~Rev.~D}%% Physical Review D 
\newcommand{\prl}{Phys.~Rev.~Lett.}%% Physical Review Letters 
\newcommand{\pasj}{PASJ}% % Publications of the ASJ 
\newcommand{\nat}{Nature}%    % Nature 
\newcommand{\procspie}{Proc.~SPIE}% 
\def\lesssim{\mathrel{\hbox{\rlap{\hbox{\lower4pt\hbox{$\sim$}}}\hbox{$<$}}}}
\def\gtrsim{\mathrel{\hbox{\rlap{\hbox{\lower4pt\hbox{$\sim$}}}\hbox{$>$}}}}

\begin{document}
\jname{To appear in Annu. Rev. Astronomy \& Astrophysics}
\jyear{2002}
\jvol{40}
\ARinfo{}
%\ARinfo{1056-8700/97/0610-00}

\input epsf.tex
\input psfig.sty

\title{Cosmology with the Sunyaev-Zel'dovich Effect}

\markboth{Carlstrom, Holder, \& Reese}{Cosmology with SZE}

\author{John E.\ Carlstrom$^{1}$, Gilbert P.\ Holder$^{2}$, and Erik D.\ Reese$^{3}$
\affiliation{$^1$Center for Cosmological Physics, Department of Astronomy and Astrophysics, Department of Physics, Enrico Fermi Institute, University of Chicago, 5640 S. Ellis Avenue, Chicago, IL, 60637, email: jc@hyde.uchicago.edu\\
$^2$Institute for Advanced Study, Princeton, NJ, 08540, email: holder@ias.edu\\
$^3$Chandra Fellow, Department of Physics, University of California, Berkeley, CA, 94720, email: reese@cfpa.berkeley.edu}}


\begin{abstract} The Sunyaev-Zel'dovich effect (SZE) provides a unique way to map the
large-scale structure of the universe as traced by massive clusters of
galaxies. As a spectral distortion of the cosmic microwave background,
the SZE is insensitive to the redshift of the galaxy cluster, making
it well-suited for studies of clusters at all redshifts, and
especially at reasonably high redshifts ($z > 1$) where the abundance
of clusters is critically dependent on the underlying cosmology.
Recent high signal-to-noise detections of the SZE have enabled
interesting constraints on the Hubble constant and the matter density
of the universe using small samples of galaxy clusters. Upcoming SZE
surveys are expected to find hundreds to thousands of new galaxy
clusters, with a mass selection function that is remarkably uniform
with redshift.  In this review we provide an overview of the SZE and
its use for cosmological studies with emphasis on the cosmology
that can, in principle, be extracted from SZE survey yields. We
discuss the observational and theoretical challenges that must be met
before precise cosmological constraints can be extracted from the
survey yields.

\end{abstract}

\maketitle

%%%%%%%%%%%%%%%%%%%%%%%%%%%%%%%%%%%%%%%%%%%%%%%%%%%%%%%%%%%%%%%%%%%%%%%%%%%%%
%%%  									  %%%
%%%	I.   Introduction						  %%%
%%%  									  %%%
%%%%%%%%%%%%%%%%%%%%%%%%%%%%%%%%%%%%%%%%%%%%%%%%%%%%%%%%%%%%%%%%%%%%%%%%%%%%%

\section{INTRODUCTION}
\label{sec:intro}

The Sunyaev-Zel'dovich Effect (SZE) offers a unique and powerful
observational tool for cosmology.  Recently, there has been
considerable progress in detecting and imaging the SZE.  Efforts over
the first two decades after the SZE was first proposed in 1970
\citep{sunyaev70,sunyaev72} yielded few reliable detections.  Over the
last decade, new detectors and observing techniques have allowed high
quality detections and images of the effect for more than 50 clusters
with redshifts as high as one.  The next generation of SZE instruments
that are now being built or planned will be orders of magnitude more
efficient. Entering the fourth decade of SZE observations, we are now
in position to exploit fully the power of the SZE, by obtaining
detailed images of a set of clusters to understand the intra-cluster
medium (ICM), by obtaining large SZE samples of clusters to determine
statistically robust estimates of the cosmological parameters and,
most importantly, by conducting large untargeted SZE surveys to probe
the high redshift universe. These surveys will provide a direct view
of the growth of large scale structure and will provide large catalogs
of clusters that extend past $z \sim 2$ with remarkably uniform
selection functions.

The physics of the SZE has been covered well in previous reviews
\citep{birkinshaw99, rephaeli95,sunyaev80}, with \citet{birkinshaw99}
and \citet{carlstrom00} providing recent reviews of the observations.
In this review, we look to the near future, using recent observations
as a guide to what we can expect.

The SZE is best known for allowing the determination of cosmological
parameters when combined with other observational diagnostics of
clusters of galaxies such as X-ray emission from the intracluster gas,
weak and strong lensing by the cluster potential, and optical galaxy
velocity dispersion measurements.  For example, cluster distances have
been determined from the analysis of SZE and X-ray data, providing
independent estimates of the Hubble constant. A large homogeneous
sample of galaxy clusters extending to high redshift should allow a
precise measure of this number, as well as a measure of the angular
diameter distance relation to high redshift where it is highly
sensitive to cosmological parameters.  Similarly, the SZE and X-ray
measurements will allow tight constraints on cluster gas mass
fractions which can be used to estimate $\Omega_M$ assuming the
composition of clusters represents a fair sample of the universal
composition. The observed redshift dependence of the gas mass fraction
can also be used to constrain cosmological parameters as well as test
speculative theories of dark matter decay.

The most unique and powerful cosmological tool provided by the
exploitation of the SZE will likely be the direct measurement of the
evolution of the number density of galaxy clusters by deep, large
scale SZE surveys.  The redshift evolution of the cluster density is
critically dependent on the underlying cosmology, and in principle can
be used to determine the equation of state of the dark energy.
SZE observations are particularly well suited for deep surveys because
the important parameter that sets the detection limit for such a
survey is the mass of the cluster;
SZE surveys will be able to detect all clusters above a mass
limit independent of the redshift of the clusters.  This remarkable
property of SZE surveys is due to the fact that the SZE is a
distortion of the cosmic microwave background (CMB) spectrum. While
the CMB suffers cosmological dimming with redshift, the ratio of the
magnitude of the SZE to the CMB does not; it is a direct, redshift
independent measurement of the ICM column density weighted by
temperature, i.e., the pressure integrated along the line of
sight. The total SZE flux detected will be proportional to the total
temperature-weighted mass (total integrated pressure) and, of course,
inversely proportional to the square of the angular diameter
distance. Adopting a reasonable cosmology and accounting for the
increase in the universal matter density with redshift, the mass limit
for a given SZE survey flux sensitivity is not expected to change more
than a factor of $\sim 2-3$ for any clusters with $z>0.05$.

SZE surveys therefore offer an ideal tool for determining the cluster
density evolution. Analyses of even a modest survey covering $\sim$~10
square degrees will provide interesting constraints on the matter
density of the universe. The precision with which cosmological
constraints can be extracted from much larger surveys, however, will
be limited by systematics due to our insufficient understanding of the
structure of clusters, their gas properties and evolution.

Insights into the structure of clusters will be provided by high
resolution SZE observations, especially when combined with other
measurements of the clusters.  Fortunately, many of the cluster
properties derived directly from observational data can be determined
in several different ways.  For example, the gas mass fraction can be
determined by various combinations of SZE, X-ray, and lensing
observations. The electron temperature, a direct measure of a
cluster's mass, can be measured directly through X-ray spectroscopy,
or determined through the analysis of various combinations of X-ray,
SZE, and lensing observations.  Several of the desired properties of
clusters are therefore over-constrained by observation, providing
critical insights to our understanding of clusters, and critical tests
of current models for the formation and evolution of galaxy clusters.
With improved sensitivity, better angular resolution, and sources out
to $z \sim 2$, the next generation of SZE observations will provide a
good view of galaxy cluster structure and evolution. This will allow,
in principle, the dependence of the cluster yields from large SZE
surveys on the underlying cosmology to be separated from the
dependence of the yields on cluster structure and evolution.

We outline the properties of the SZE in the next section and provide
an overview of the current state of the observations in
\S\ref{sec:obs_status}.  This is followed in \S\ref{sec:survey_yields}
by predictions for the expected yields of upcoming SZE surveys. In
\S\ref{sec:sze_cosmo}, we provide an overview of the cosmological
tests which will be possible with catalogs of SZE-selected
clusters. This is followed by a discussion of backgrounds,
foregrounds, contaminants, and theoretical uncertainties that could
adversely affect cosmological studies with the SZE and a discussion of
observations which could reduce or eliminate these concerns.
Throughout the paper, $h$ is used to parametrize the Hubble constant
by $H_0 = 100 h$ km s$^{-1}$ Mpc$^{-1}$, and $\OmM$ and $\OmL$ are the
matter density and vacuum energy density, respectively, in units of
the critical density.

%%%%%%%%%%%%%%%%%%%%%%%%%%%%%%%%%%%%%%%%%%%%%%%%%%%%%%%%%%%%%%%%%%%%%%%%%%%%%
%%%  									  %%%
%%%	II.  The Sunyaev-Zel'dovich Effect				  %%%
%%%  									  %%%
%%%%%%%%%%%%%%%%%%%%%%%%%%%%%%%%%%%%%%%%%%%%%%%%%%%%%%%%%%%%%%%%%%%%%%%%%%%%%

\section{THE SUNYAEV-ZEL'DOVICH EFFECT}
\label{sec:sze}

\begin{figure}[!tbh]
\centerline{\psfig{figure=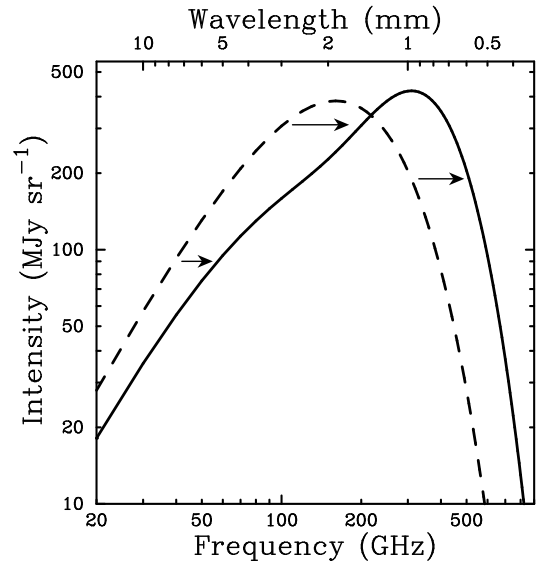,height=3.5in}}
\caption{The Cosmic Microwave Background (CMB) spectrum, undistorted
(dashed line) and distorted by the Sunyaev-Zel'dovich effect (SZE)
(solid line).  Following \protect\citet{sunyaev80} to illustrate the effect,
the SZE distortion shown is for a fictional cluster 1000 times more
massive than a typical massive galaxy cluster.  The SZE causes a
decrease in the CMB intensity at frequencies $\lsim 218$ GHz and an
increase at higher frequencies.}
\label{fig:sze_cmb}
\end{figure}

\begin{figure}[!tbh]
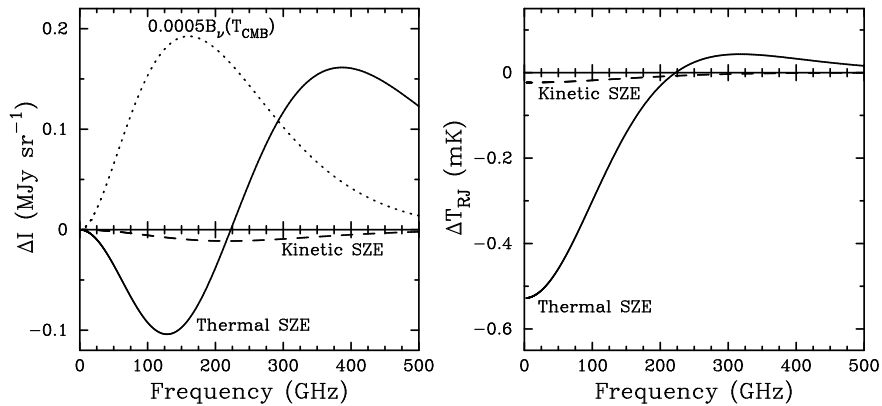

\centerline{
	\psfig{figure=./dI_araa.ps,height=2.5in}
	\psfig{figure=./dT_araa.ps,height=2.5in}
}
\caption{Spectral distortion of the cosmic microwave background (CMB)
radiation due to the Sunyaev-Zel'dovich effect (SZE). The left panel
shows the intensity and the right panel shows the Rayleigh Jeans
brightness temperature. The thick solid line is the thermal SZE and
the dashed line is the kinetic SZE. For reference the 2.7~K thermal
spectrum for the CMB intensity scaled by 0.0005 is shown by the dotted
line in the left panel.  The cluster properties used to calculate the
spectra are an electron temperature of 10~keV, a Compton $y$~parameter
of $10^{-4}$, and a peculiar velocity of 500 km s$^{-1}$.}
\label{fig:spectrum}
\end{figure}

\subsection{Thermal Sunyaev-Zel'dovich Effect}
\label{subsec:thermal_sze}

The Sunyaev-Zel'dovich effect (\sze) is a small spectral distortion of
the cosmic microwave background (CMB) spectrum caused by the
scattering of the CMB photons off a distribution of high energy
electrons. Here we focus only on the SZE caused by the hot thermal
distribution of electrons provided by the intra-cluster medium (ICM)
of galaxy clusters.  CMB photons passing through the center of a
massive cluster have only a $\approx 1$\% probability of
interacting with an energetic ICM electron.  The resulting inverse
Compton scattering preferentially boosts the energy of the CMB photon
by roughly $\kb T_e / m_e c^2$ causing a small ($\lsim 1$ mK)
distortion in the CMB spectrum.  Figure~\ref{fig:sze_cmb} shows the
SZE spectral distortion for a fictional cluster that is over 1000
times more massive than a typical cluster to illustrate the small
effect.  The SZE appears as a decrease in the intensity of the CMB at
frequencies $\lsim 218$ GHz and as an increase at higher
frequencies.

The derivation of the \sze\ can be found in the original papers of
Sunyaev and Zel'dovich \citep{sunyaev70, sunyaev72}, in several
reviews \citep{sunyaev80, rephaeli95, birkinshaw99}, and in a number
of more recent contributions which include relativistic corrections
(see below for references).  This review discusses the basic features of the SZE that
make it a useful cosmological tool.
% are discussed.

The SZE spectral distortion of the CMB expressed as a temperature
change $\Delta T_{SZE}$ at dimensionless frequency $x \equiv
\frac{h\nu}{k_BT_{CMB}}$ is given by
\begin{equation}
   \frac{\Delta T_{SZE}}{T_{CMB}} =  f(x) \ y  = f(x) \int
   n_e \frac{k_B T_e}{m_e c^2} \sigma_T \, d\ell, 
   \label{eq:deltaT1}\label{eq:y}
\end{equation}
where $y$ is the Compton $y$-parameter, which for an isothermal
cluster equals the optical depth, $\tau_e$, times the fractional energy gain per
scattering, $\sigma_T$ is the Thomson cross-section, $n_e$ is the
electron number density, $T_e$ is the electron temperature, $k_B$ is
the Boltzmann constant, $m_e c^2$ is the electron rest mass energy,
and the integration is along the line of sight.  The frequency
dependence of the \sze\ is
\begin{equation}
   f(x) = \left(x \frac{e^x+1}{e^x-1} -4\right)(1 + \delta_{SZE}(x,T_e)),
   \label{eq:fx}
\end{equation}
where $\delta_{SZE}(x,T_e)$ is the relativistic correction to the
frequency dependence.  Note that $f(x) \rightarrow -2$ in the
non-relativistic and Rayleigh-Jeans (RJ) limits.

It is worth noting that $\Delta T_{SZE} / \Tcmb$ is independent of
redshift as shown in Eq.~\ref{eq:deltaT1}. This unique feature of the
SZE makes it a potentially powerful tool for investigating the high
redshift universe.

Expressed in units of specific intensity, common in millimeter SZE
observations, the thermal SZE is
\begin{equation}
\Delta I_{SZE} = g(x) I_0 y,
\label{eq:sze_intensity}
\end{equation}
where $I_0 = 2 (\kB\Tcmb)^3 / (h c)^2$ and the frequency dependence is
given by
\begin{equation}
g(x) = \frac{x^4 e^x}{(e^x-1)^2} \left ( x \frac{e^x + 1}{e^x - 1} - 4
\right ) \left ( 1 + \delta_{SZE}(x, T_e) \right ).
\label{eq:gx}
\end{equation}
$\Delta T_{SZE}$ and $\Delta I_{SZE}$ are simply related by the
derivative of the blackbody with respect to temperature, $\left |
dB_\nu / dT \right |$.

The spectral distortion of the CMB spectrum by the thermal SZE is
shown in Figure~\ref{fig:spectrum} (solid line) for a realistic
massive cluster ($y = 10^{-4}$) in units of intensity (left panel) and
Rayleigh-Jeans (RJ) brightness temperature (right panel).  The RJ
brightness is shown because the sensitivity of a radio telescope is
calibrated in these units. It is defined simply by $I_\nu = (2 k_B
\nu^2/c^2) T_{RJ}$ where $I_\nu$ is the intensity at frequency $\nu$,
$k_B$ is Boltzmann's constant, and $c$ is the speed of light.  The CMB
blackbody spectrum, $B_\nu (\Tcmb)$, multiplied by 0.0005 (dotted
line) is also shown for comparison.  Note that the spectral signature
of the thermal effect is distinguished readily from a simple
temperature fluctuation of the CMB.  The kinetic SZE distortion is
shown by the dashed curve (\S\ref{sec:kinetic_sze}).  In the
non-relativistic regime, it is indistinguishable from a CMB
temperature fluctuation.

The gas temperatures measured in massive galaxy clusters are around
$k_B T_e \sim 10$ keV \citep{mushotzky97, allen98} and are measured to
be as high as $\sim 17$ keV in the galaxy cluster 1E $0657-56$
\citep{tucker98}. The mass is expected to scale with temperature
roughly as $T_e \propto M^{2/3}$.  At these temperatures, electron
velocities are becoming relativistic and small corrections are
required for accurate interpretation of the SZE.  There has been
considerable theoretical work to include relativistic corrections to
the SZE \citep{wright79, fabbri81, rephaeli95, rephaeli97, stebbins97,
itoh98, challinor98, sazonov98, sazonov98b, nozawa98, challinor99,
molnar99, dolgov01}.  All of these derivations agree for $\kb T_e
\lsim 15$ keV, appropriate for galaxy clusters.  For a massive cluster
with $k_BT_e \sim 10$keV ($k_BT_e/m_ec^2 \sim 0.02$) the relativistic
corrections to the SZE are of order a few percent in the RJ portion of
the spectrum, but can be substantial near the null of the thermal
effect. Convenient analytical approximations to fifth order in
$k_BT_e/m_e c^2$ are presented in \citet{itoh98}.

Particularly relevant for finding clusters with an SZE survey is the
integrated \sze\ signal.  Because the SZE signal is the integrated
pressure, integrating over the solid angle of the cluster provides a
sum of all of the electrons in the cluster weighted by temperature.
This provides a relatively clean measure of the total thermal energy
of the cluster.  Integrating the SZE over the solid angle of the
cluster, $d\Omega = dA / \Da^{\!\!\!2}$, gives
\begin{equation}
\int \Delta T_{SZE} \; d\Omega \propto \frac{N_e \left < T_e \right
>}{\Da^{\!\!\!2}} \propto \frac{M \left < T_e \right
>}{\Da^{\!\!\!2}} 
\label{eq:tot_sze_flux}
\end{equation}
where $N_e$ is the total number of electrons in the clusters, $\left <
T_e \right >$ is the mean electron temperature, $\Da$ is the angular
diameter distance, and $M$ is the mass of the cluster (either gas or
total mass as $M_{gas} = M_{total}f_g$, where $f_g$ is the gas mass
fraction).  The integrated SZE flux is simply the temperature weighted
mass of the cluster divided by $\Da^{\!\!\!2}$. The angular diameter
distance $\Da(z)$ is fairly flat at high redshift. Also, a cluster of
a given mass will be denser and therefore hotter at high redshift
because the universal matter density increases as $(1 + z)^3$.
Therefore, one expects an SZE survey to detect all clusters above some
mass threshold with little dependence on redshift (see
\S\ref{subsec:mass_limit}).

The most important features of the thermal \sze\ are: 1) it is a small
spectral distortion of the CMB of order $\sim 1$ mK, which is
proportional to the cluster pressure integrated along the line of
sight [Eq.~\ref{eq:y}]; 2) it is independent of redshift; 3) it has a
unique spectral signature with a decrease in the CMB intensity at
frequencies $\lsim 218$ GHz and an increase at higher frequencies; and
4) the integrated \sze\ flux is proportional to the temperature
weighted mass of the cluster (total thermal energy of the cluster)
implying that SZE surveys will have a mass threshold nearly
independent of redshift.

\subsection{Kinetic Sunyaev-Zel'dovich Effect}
\label{sec:kinetic_sze}

If the cluster is moving with respect to the CMB rest frame there will
be an additional spectral distortion due to the Doppler effect of the
cluster bulk velocity on the scattered CMB photons.  If a component of
the cluster velocity, $v_{pec}$, is projected along the line of sight
to the cluster, then the Doppler effect will lead to an observed
distortion of the CMB spectrum referred to as the kinetic SZE. In the
non-relativistic limit, the spectral signature of the kinetic SZE is a
pure thermal distortion of magnitude
\begin{equation}
\frac{\Delta T_{SZE}}{T_{CMB}}  = -\tau_e \left ( \frac{v_{pec}}{c} \right ),
\label{eq:v_pec}
\end{equation}
where $v_{pec}$ is along the line of sight; i.e., the emergent
spectrum is still described completely by a Planck spectrum, but at a
slightly different temperature, lower (higher) for positive (negative)
peculiar velocities (\citealt{sunyaev72, phillips95, birkinshaw99},
see Figure~\ref{fig:spectrum}).

Relativistic perturbations to the kinetic SZE are due to the Lorentz
boost to the electrons provided by the bulk velocity \citep{nozawa98b,
sazonov98}. The leading term is of order $(k_B T_e/m_e
c^2)(v_{pec}/c)$ and for a 10~keV cluster moving at 1000 \kms\ the
effect is about an 8\% correction to the non-relativistic term. The
$(k_B T_e/m_e c^2)^2(v_{pec}/c)$ term is only about 1\% of the
non-relativistic kinetic \sze\ and the $(v_{pec}/c)^2$ term is only
0.2\%.

\subsection{Polarization of the Sunyaev-Zel'dovich Effect}
\label{subsec:polar_sze}

The scattering of the CMB photons by the hot ICM electrons can result
in polarization at levels proportional to powers of $(v_{pec}/c)$ and
$\tau_e$.  The largest polarization is expected from the anisotropic
optical depth to a given location in the cluster. For example, toward
the outskirts of a cluster one expects to see a concentric (radial)
pattern of the linear polarization at frequencies where the thermal
SZE is positive (negative). Plots of the polarization pattern are
presented in \citet{sazonov99}. Nonspherical morphology for the
electron distributions will lead to considerably complicated
polarization patterns.  The peak polarization of this signal will be
order $\tau_e$ times the SZE signal, i.e., of order $0.025 (\kb
T_e/m_ec^2)\tau_e^2$ times the CMB intensity. For a massive cluster
with $\tau_e = 0.01$, the effect would be at the 0.1~$\mu$K level
toward the edge of the cluster. In principle, this effect could be
used to measure the optical depth of the cluster and therefore
separate $T_e$ and $\tau_e$ from a measurement of the thermal \sze\
(see Eq.~\ref{eq:deltaT1}).

It can be shown that polarization of the \sze\ comes entirely from the
quadrupole component of the local radiation field experienced by the
scattering electron. In the case above, the quadrupole component at
the outskirts of the cluster is due to the anisotropy in the radiation
field in the direction of the cluster center due to the \sze.  Sunyaev
and Zel'dovich discussed polarization due to the motion of the cluster
with respect to the CMB and transverse to our line of sight
(\citealt{sunyaev80b}, see also \citealt{sazonov99}). In this case,
the quadrupole comes from the Doppler shift. They found the largest
terms to be of order $0.1\tau_e (v_{pec}/c)^2$ and
$0.025\tau_e^2(v_{pec}/c)$ of the CMB intensity. The latter term,
second order in $\tau_e$, can be thought of as imperfect cancellation
of the dipole term due to the anisotropic optical depth.  Using
$\tau_e = 0.01$ and a bulk motion of 500 km s$^{-1}$, results in
polarization levels of order $10~nK$, far beyond the sensitivity of
current instrumentation.

The CMB as seen by the cluster electrons will have a quadrupole
component and therefore the electron scattering will lead to linear
polarization.  This mechanism could possibly be used to trace the
evolution of the CMB quadrupole if polarization measurements could be
obtained for a large number of clusters binned in direction and
redshift \citep{kamionkowski97a,sazonov99}. Sazonov and Sunyaev
calculated the expected polarization level and found the maximum CMB
quadrupole induced polarization is $50 (\tau_e/0.01)$~nK, somewhat
higher than the expected velocity induced terms discussed above.  The
effect is again too small to expect detection in the near future.
However, by averaging over many clusters, detecting this polarization
might be possible with future satellite missions.

%%%%%%%%%%%%%%%%%%%%%%%%%%%%%%%%%%%%%%%%%%%%%%%%%%%%%%%%%%%%%%%%%%%%%%%%%%%%%
%%%  									  %%%
%%%	III. Current status of observations				  %%%
%%%  									  %%%
%%%%%%%%%%%%%%%%%%%%%%%%%%%%%%%%%%%%%%%%%%%%%%%%%%%%%%%%%%%%%%%%%%%%%%%%%%%%%

\section{STATUS OF OBSERVATIONS}
\label{sec:obs_status}

In the twenty years following the first papers by Sunyaev and
Zel'dovich \citep{sunyaev70, sunyaev72} there were few firm detections
of the \sze\ despite a considerable amount of effort
\citep{birkinshaw91}. Over the last several years, however,
observations of the effect have progressed from low S/N detections and
upper limits to high confidence detections and detailed images. In
this section we briefly review the current state of \sze\
observations.

The dramatic increase in the quality of the observations is due to
improvements both in low-noise detection systems and in observing
techniques, usually using specialized instrumentation to control
carefully the systematics that often prevent one from obtaining the
required sensitivity.  The sensitivity of a low-noise radio receiver
available 20 years ago should have easily allowed the detection of the
SZE toward a massive cluster. Most attempts, however, failed due to
uncontrolled systematics.  Now that the sensitivities of detector
systems have improved by factors of 3 to 10, it is clear that the goal
of all modern SZE instruments is the control of systematics.  Such
systematics include, for example, the spatial and temporal variations
in the emission from the atmosphere and the surrounding ground, as
well as gain instabilities inherent to the detector system used.

The observations must be conducted on the appropriate angular scales.
Galaxy clusters have a characteristic scale size of order a
megaparsec. For a reasonable cosmology, a megaparsec subtends an
arcminute or more at any redshift; low redshift clusters will subtend
a much larger angle, for example the angular extent of the Coma
cluster ($z = 0.024$) is of order a degree (core radius $\sim 10'$)
\citep{herbig95}.  The detection of extended low surface brightness
objects requires precise differential measurements made toward widely
separated directions on the sky. The large angular scale presents
challenges to control offsets due to differential ground pick-up and
atmospheric variations.

\subsection{Sources of Astronomical Contamination and Confusion}
\label{sec:confusion}

\begin{figure}[!tbh]
\centerline{
   \psfig{figure=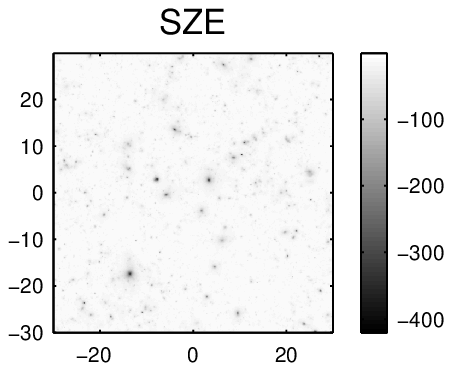,width=0.33\columnwidth}
   \psfig{figure=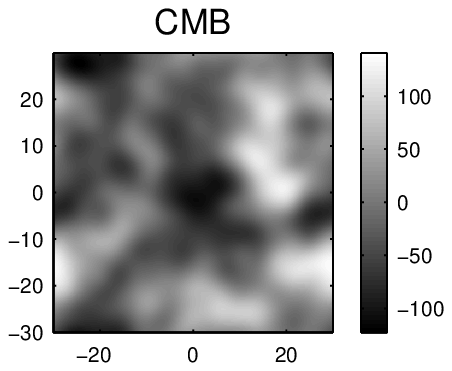,width=0.33\columnwidth}
   \psfig{figure=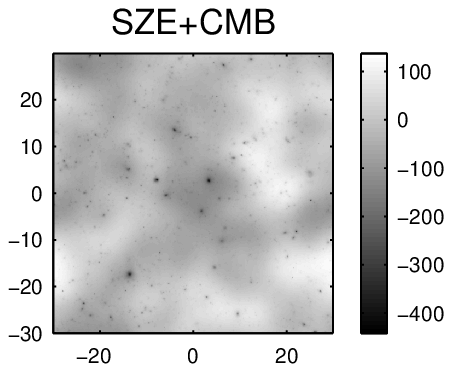,width=0.33\columnwidth}
}
\caption{Illustration of the characteristic angular scales of primary
CMB anisotropy and of the \sze.  The images each cover one square
degree and the gray scales are in $\mu$K. The left panel shows a image
of the SZE from many clusters at $150\,$GHz (2 mm) obtained from a state-of-the-art hydrodynamic
simulation \protect\citep{springel00}. The clusters appear point-like at
this angular scale. The center panel is a realization of
CMB anisotopy for a $\Lambda$CDM cosmology.  The right panel
illustrates that the \sze\ can be distinguished readily from primary
CMB anisotropy provided the observations have sufficient angular
resolution.
\label{fig:cmb-sze}
}
\end{figure}

In designing an instrument for SZE observation, one also needs to take
into account several sources of possible contamination and confusion
from astronomical sources. One such source is anisotropy of the CMB
itself (see Figure~\ref{fig:cmb-sze}). For distant clusters with
angular extents of a few arcminutes or less it is not a serious
problem as the CMB anisotropy is expected \citep{hu97} and indeed
found to be damped considerably on these scales
(\citealt{church97,subrahmanyan00,dawson01}, see also
\citealt{holzapfel97b} and \citealt{laroque01} for CMB limits to \sze\
contamination).  For nearby clusters, or for searches for distant
clusters using beams larger than a few arminutes, the intrinsic CMB
anisotropy must be considered. The unique spectral behavior of the
thermal \sze\ can be used to separate it from the intrinsic CMB in
these cases. Note, however, that for such cases it will not be
possible to separate the kinetic \sze\ effects from the intrinsic CMB
anisotropy without relying on the very small spectral distortions of
the kinetic \sze\ due to relativistic effects.

Historically, the major source of contamination in the measurement of
the \sze\ has been radio point sources. It is obvious that emission
from point sources located along the line of the sight to the cluster
could fill in the \sze\ decrement, leading to an underestimate. The
radio point sources are variable and therefore must be
monitored. Radio emission from the cluster member galaxies, from the
central cD galaxy in particular, is often the largest source of radio
point source contamination, at least at high radio frequencies
\citep{cooray98, laroque01}.  The typical spectral index of the radio
point sources is $\alpha \sim 0.7$ for $S_\nu \propto \nu^{-\alpha}$,
where $S_\nu$ is the point source flux. In the RJ limit, the \sze\
flux is proportional to $\nu^2$ and therefore point sources are much
less of an issue at higher radio frequencies.

Although it is most likely that insufficient attention to radio point
sources would lead to the underestimate of the \sze\ effect, it could
also lead to an overestimate. The most obvious example is if
unaccounted point sources are in the reference fields surrounding the
cluster.  An effect due to gravitational lensing has also been pointed
out for low frequency observations where the flux from many point
sources must be taken into account before a reliable measure of the
\sze\ can be made.  Essentially, gravitational lensing increases the 
efficiency of detecting
point sources toward the center of the cluster 
which could lead to an overestimate of the \sze\ decrement
\citep{loeb1997}. This effect should be negligible at frequencies
greater than roughly 30~GHz.

At frequencies near the null of the thermal \sze\ and higher, dust
emission from extragalactic sources as well as dust emission from our
own galaxy must be considered. Dust emission from our Galaxy rises
steeply as $\nu^{2+\beta}$ with the observed dust opacity index
$\beta$ found to be $0 < \beta < 2 $ over the frequencies of interest.

At the angular scales and frequencies of interest for most SZE
observations, contamination from diffuse Galactic dust emission will
not usually be significant and is easily compensated.  Consider
instead the dusty extragalactic sources such as those that have been
found toward massive galaxy clusters with the SCUBA bolometer array
\citep{smail97}. Spectral indices for these sources are estimated to
be $\sim 1.5-2.5$ \citep{blain98,fischer93}.  Sources with 350~GHz
(850~$\mu m$) fluxes greater than 8~mJy are common and all clusters
surveyed had multiple sources with fluxes greater than 5~mJy. A 10~mJy
source at 350~GHz corresponds to $\Delta T_{CMB} = 345\ \mu$K for 1$'$
beam, or a Compton $y$-parameter of $6\times 10^{-5}$.  The same
source scaled to 270~GHz, assuming a $\nu^{2}$ spectrum, corresponds
to $\Delta T_{CMB} = 140\ \mu$K at 270~GHz for 1$'$ beam and a
$y$-parameter of $6\times 10^{-5}$. Scaling to the SZE thermal null at
218 GHz gives 3.9 mJy which corresponds to a $\Delta T_{CMB} = 85\
\mu$K for a 1$'$ beam. This in turn translates directly to an
uncertainty in a measurement of the cluster peculiar velocity
(Eq.~\ref{eq:v_pec}); for a massive cluster with an optical depth of
0.01 and an electron temperature of 10~keV, 85~$\mu$K corresponds to
a peculiar velocity of 930 \kms.  The contamination is more severe for
less massive clusters with the dependence scaling as $\Delta v_{pec}
\propto \tau_e^{-1} \propto R^2/M \propto M^{-1/3} \propto
T_e^{-1/2}$. The contamination scales inversely with the beam area.

As with SZE observations at radio frequencies, the analyses of 
high frequency observations
also need to consider the effects of point sources and require
either high dynamic angular range, large spectral coverage, or both,
to separate the point source emission from the SZE.

\subsection{Single Dish Observations}
\label{sec:single_dish}

All observations sensitive enough to observe the \sze\ are
differential.  The primary issue for single dish observations is how
to switch the beam on the sky without introducing systematics
comparable to the SZE.  This beam switching can be accomplished in
several ways, including but not limited to Dicke switching between
feeds and chopping mirrors which switch or sweep the beam on the sky.
With a single dish telescope, modulation of the beam sidelobes can
lead to an offset.  This offset can be removed if it remains stable
enough to be measured on some portion of the sky without a
cluster.  However, temperature variations of the optics and features
on the ground will cause the offset to change as the source is
tracked.  Therefore, it has become common practice to observe leading
and trailing fields when they have the same position with respect to
the ground as the source.  In this way, any constant or linear drift
in offset can be removed at the price of observing efficiency and
sensitivity.

The first measurements of the \sze\ were made with single dish radio
telescopes at centimeter wavelengths.  Advances in detector technology
made the measurements possible, although early observations appear to
have been plagued by systematic errors that led to irreproducible and
inconsistent results.  Eventually, successful detections using beam
switching techniques were obtained.  During this period, the
pioneering work of Birkinshaw and collaborators with the OVRO 40 meter
telescope stands out for its production of results which served to
build confidence in the technique \citep{birkinshaw78a,
birkinshaw78b,birkinshaw91}.  More recently, leading and trailing beam
switching techniques have been used successfully with the OVRO 5 meter
telescope at 32 GHz to produce reliable detections of the \sze\ in
several intermediate redshift clusters \citep{herbig95, myers97,
mason01}. The SEST 15 meter and IRAM 30 meter telescopes have been
used with bolometric detectors at 140 GHz and chopping mirrors to make
significant detections of the SZE effect in several clusters
\citep{andreani96, andreani99, desert98, pointecouteau99,
pointecouteau01}.  The Nobeyama 45~m telescope has also been been used
at 21~GHz, 43~GHz, and 150~GHz to detect and map the SZE
\citep{komatsu01, komatsu99}.

In the Sunyaev-Zel'dovich Infrared Experiment (SuZIE), pixels in a
six element 140 GHz bolometer array are electronically differenced by
reading them out in a differential bridge circuit
\citep{holzapfel97}. Differencing in this way makes the experiment
insensitive to temperature and amplifier gain fluctuations that
produce 1/f noise.  This increased low frequency stability allows
SuZIE to observe in a drift scanning mode where the telescope is fixed
and the rotation of the earth moves the beams across the sky.  Using
this drift scanning technique, the SuZIE experiment has produced high
signal to noise strip maps of the SZE emission in several clusters
\citep{holzapfel97b,mauskopf00}.

\begin{figure}[!tbh]
\centerline{\psfig{figure=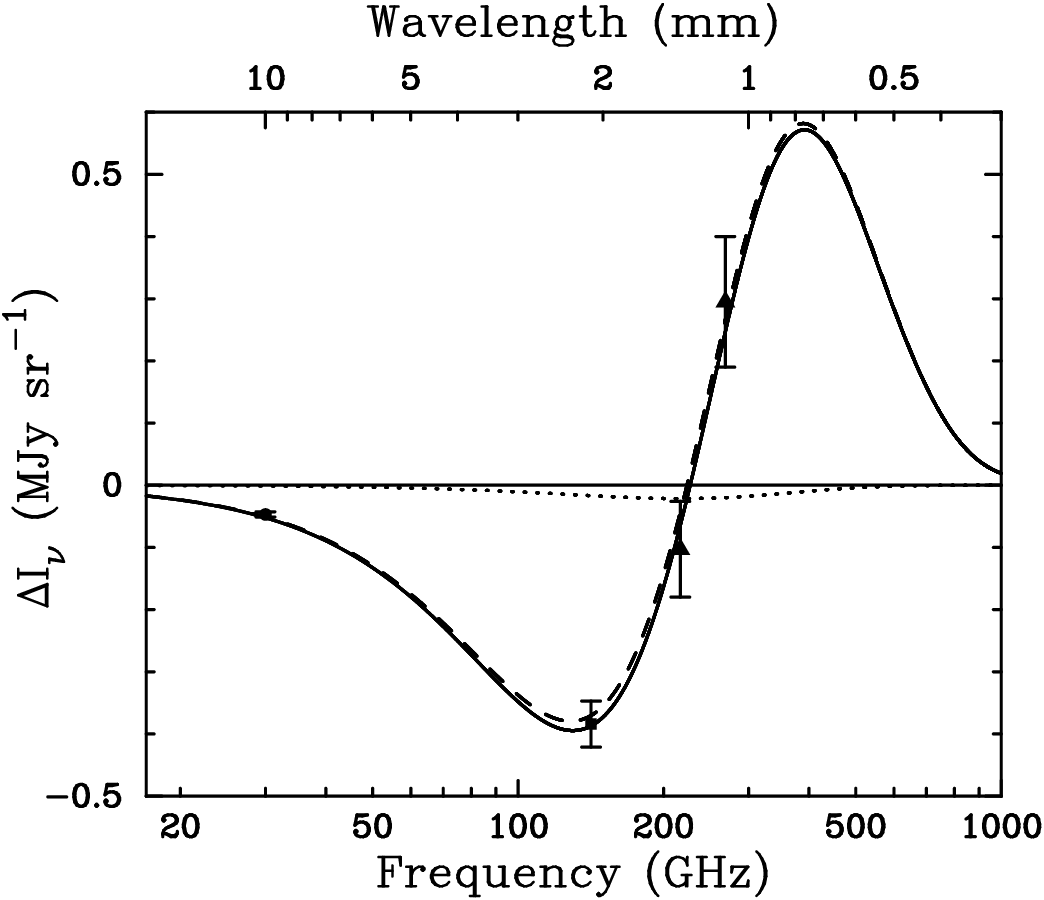,height=3.0in}}
\caption{The measured SZE spectrum of Abell 2163. The data point at 30
GHz is from the Berkeley-Illinois-Marlyand-Association (BIMA) array
\protect\citep{laroque01}, at 140~GHz it is the weighted average of Diabolo
and SuZIE measurements \protect\citep{desert98,holzapfel97b} 
and at 218~GHz and 270~GHz from SuZIE \protect\citep{holzapfel97b}. 
The best fit thermal and kinetic SZE spectra are shown by
the dashed line and the dotted lines, respectively, with the spectra
of the combined effect shown by the solid line. The limits on the
Compton $y$-parameter and the peculiar velocity are
$y_0=3.56^{+0.41}_{-0.41}\, ^{+0.27}_{-0.19}\times10^{-4}$ and
$v_p=410^{+1030}_{-850}\, ^{+460}_{-440}$ km s$^{-1}$,
respectively, with statistical followed by systematic uncertainties at
68\% confidence \protect\citep{holzapfel97b,laroque01}. }
\label{fig:a2163_spec}
\end{figure}

Because of the high sensitivity of bolometric detectors at millimeter
wavelengths, single dish experiments are ideally suited for the
measurement of the SZE spectrum.  By observing at several millimeter
frequencies these instruments should be able to separate the thermal
and kinetic SZEs from atmospheric fluctuations and sources of
astrophysical confusion.  One of the first steps to realizing this
goal is the measurement of SZE as an increment.  So far, there have
been only a few low signal to noise detections at a frequency of
approximately 270 GHz.  The main reason for the lack of detection is
the increased opacity of the atmosphere at higher frequencies.
\citet{holzapfel97b} report a detection of Abell 2163 with the SuZIE
instrument at 270 GHz.  \citet{andreani96} claim detections of the SZE
increment in two clusters, although observations of a third cluster
appear to be contaminated by foreground sources or systematic errors
\citep{andreani99}.

Figure~\ref{fig:a2163_spec} shows the measured SZE spectrum of Abell
2163, spanning the decrement and increment with data obtained from
different telescopes and techniques
\citep{holzapfel97b,desert98,laroque01}.  The SZE spectrum is a good
fit to the data demonstrating the consistency and robustness of modern
SZE measurements.

Single dish observations of the \sze\ are just beginning to reach
their potential and the future is very promising.  The development of
large format millimeter wavelength bolometer arrays will increase the
mapping speed of current SZE experiments by orders of magnitude.  The
first of this new generation of instruments is the BOLOCAM 151 element
bolometer array \citep{mauskopf00b, glenn98} which will soon begin
routine observations at the Caltech Submillimeter Observatory.
BOLOCAM will observe in drift scanning mode and produce differences
between bolometer signals in software.  To the extent that atmospheric
fluctuations are common across a bolometric array, it will be possible
to realize the intrinsic sensitivity of the detectors.  Operating from
high astronomical sites with stable atmospheres and exceptionally low
precipitable water vapor, future large format bolometer arrays have
the potential to produce high signal to noise SZE images and search
for distant SZE clusters with unprecedented speed.

\subsection{Interferometric Observations}
\label{sec:interferometer}

The stability and spatial filtering inherent to interferometry has
been exploited to make high quality images of the \sze.  The stability
of an interferometer is due to its ability to perform simultaneous
differential sky measurements over well defined spatial frequencies.

An interferometer measures the time averaged correlation of the
signals received by a pair of telescopes -- all interferometric arrays
can be thought of as a collection of $n(n-1)/2$ two-element
interferometers.
For each pair of telescopes, the interferometer effectively multiplies
the sky brightness 
at the observing frequency by a cosine, integrates the product and
outputs the time average amplitude of the product
\citep[see][]{thompson01}.  In practice the signals are split and two
correlations are performed with one being performed with a 90 degree
relative phase shift so that the output of the interferometer,
referred to as the visibility, is the complex Fourier transform
(amplitude and phase) of the sky brightness.
The interferometer is therefore only sensitive to angular scales
(spatial frequencies) near $B/\lambda$, where the baseline $B$ is the
projected separation of the telescopes as seen by the source and
$\lambda$ is the observation wavelength.  The interferometer response
is essentially insensitive to gradients in the atmospheric emission or
other large scale emission features.

There are several other features which allow an interferometer to
achieve extremely low systematics. For example, only signals which
correlate between array elements will lead to detected signal. For
most interferometers, this means that the bulk of the sky noise for
each element will not lead to signal. Amplifier gain instabilities for
an interferometer will not lead to large offsets or false detections,
although if severe they may lead to somewhat noisy signal
amplitude. To remove the effects of offsets or drifts in the
electronics as well as the correlation of spurious (non-celestial)
sources of noise, the phase of the signal received at each telescope
is modulated and then the proper demodulation is applied to the output
of the correlator.

The spatial filtering of an interferometer also allows the emission
from radio point sources to be separated from the \sze\ emission. This
is possible because at high angular resolution ($\lsim 10''$) the
\sze\ contributes very little flux.  This allows one to use long
baselines -- which give high angular resolution -- to detect and
monitor the flux of radio point sources while using short baselines to
measure the \sze.  Nearly simultaneous monitoring of the point sources
is important as they are often time variable.  The signal from the
point sources is then easily removed, to the limit of the dynamic
range of the instrument, from the short baseline data which are
sensitive also to the \sze.

\begin{figure}[!tbh]
\centerline{
   \psfig{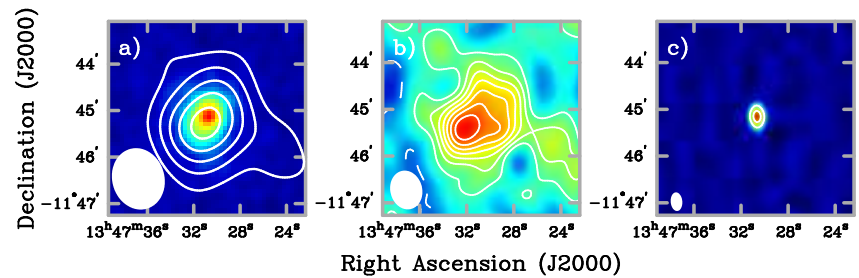}
}
\caption{
Interferometric images of the SZE of galaxy cluster
RXJ~1347-1145 emphasizing different spatial scales.  The FWHM ellipse of
the synthesized beam is shown in the lower left corner of each panel.
(a) Point source subtracted SZE image (contours) overlaid on ROSAT
X-ray image (false color).  The contours are multiples of 185$\mu$K
($\sim 2\sigma$), and negative contours are shown as solid lines.
A 1500 $\lambda$ half-power radius Gaussian taper was applied 
to the $u,v$ data resulting in a
$63'' \times 80''$ synthesized beam.  The X-ray image is HRI raw
counts smoothed with a Gaussian with $\sigma = 6''$ and contains
roughly 4000 cluster counts.  (b) Higher resolution point source
subtracted SZE image (both contours and false color).  A
3000 $\lambda$ half-power radius Gaussian taper was applied resulting 
in a $40'' \times 50''$ synthesized beam. 
The contours are multiples of 175$\mu$K
($\sim 1\sigma$) (c) Image of the point source made using projected
baselines greater than 3000 $\lambda$.  This map has a synthesized
beam of $15'' \times 24''$ and a rms of $\sim 275\ \mu$Jy beam$^{-1}$,
corresponding to a $\sim 1200\mu$K brightness sensitivity.  The
contours are multiples of 15$\sigma$. The data was taken with
the BIMA mm-array operating at 28.5 GHz. Single dish maps
of the \sze\ toward RXJ~1347-1145 have also been made with the
IRAM 30-m and the Nobeyama 45-m telescopes \protect\citep{pointecouteau01,
komatsu01}.
\label{fig:rxj1347}
}
\end{figure}

Figure~\ref{fig:rxj1347} illustrates the spatial filtering of an
interferometer with data on the galaxy cluster RXJ $1347-1145$ from
the BIMA interferometer outfitted with 30 GHz receivers
\citep[e.g.,][]{carlstrom96, carlstrom00}.  Panel $a$) shows the point
source subtracted SZE image (contours) overlaid on a ROSAT X-ray image
(color scale).  Spatial scales typical of galaxy clusters were
stressed before deconvolution.  A higher resolution SZE image is shown
in panel $b$), showing the range of spatial scales measured
by the interferometer.  Panel $c$) shows the long baseline (high
angular resolution) data only.  The bright on-center point source is
readily apparent.

For the reasons given above, interferometers offer an ideal way to
achieve high brightness sensitivity for extended low-surface
brightness emission, at least at radio wavelengths.  Most
interferometers, however, were not designed for imaging low-surface
brightness sources.  Interferometers have been built traditionally to
obtain high angular resolution and thus have employed large individual
elements for maximum sensitivity to small scale emission.  As a result
special purpose interferometric systems have been built for imaging
the SZE \citep{jones93,carlstrom96,padin01}.  All of them have taken
advantage of low-noise high-electron-mobility-transistor (HEMT) amplifiers \citep{pospieszalski95} to
achieve high sensitivity.

The first interferometric detection \citep{jones93} of the SZE was
obtained with the Ryle Telescope (RT).  The RT was built from the 5
Kilometer Array, consisting of eight 13 m telescopes located in
Cambridge, England operating at 15~GHz with East-West
configurations. Five of the telescopes can be used in a compact E-W
configuration for imaging of the \sze\ \citep{jones93, grainge93,
grainge96, grainge00, saunders00, grainger01, grainge01, jones01}.

\begin{figure}[!tbh]
\centerline{\psfig{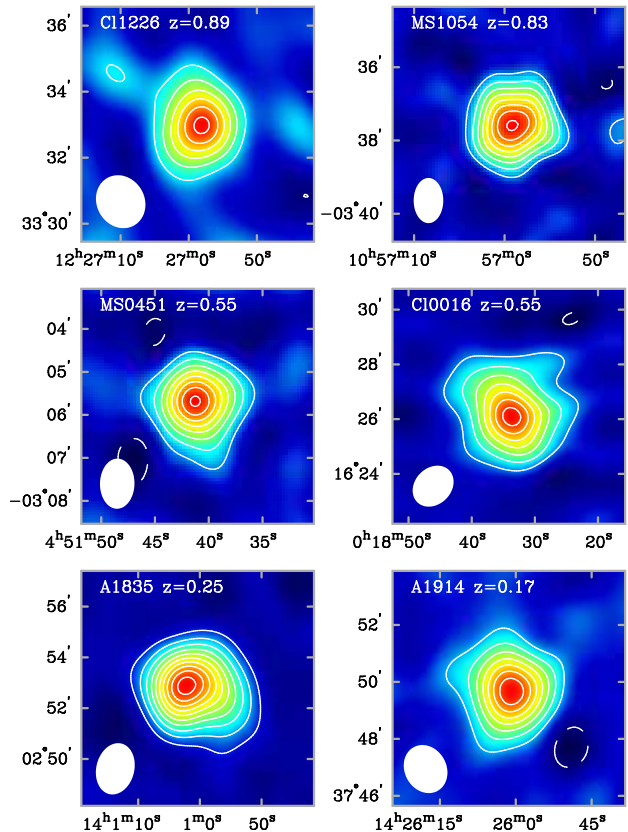}}
\caption{Deconvolved interferometric SZE images for a sample of galaxy
clusters over a large redshift range ($0.17 \leq z \leq 0.89$).  The
contours are multiples of 2$\sigma$ and negative contours are shown as
solid lines.  The FWHM ellipse of the synthesized beam is shown in the
lower left corner of each panel. The noise level $\sigma$ ranges from
$25\mu$K to $70\mu$K for the clusters shown. Radio point sources were
removed from a large fraction of the images shown.  The interferometer
was able to separate the point source emission from the SZE by using
the high resolution data obtained with long baselines. All of the
clusters shown have similar high X-ray luminosities and, as can be
seen, the strength of the \sze\ signals are similar despite the factor
of five in redshift, illustrating the independence of the \sze\ on
redshift.  }
\label{fig:szpanel12}
\end{figure}

The OVRO and BIMA SZE imaging project uses 30~GHz (1 cm) low noise
receivers mounted on the OVRO\footnote{An array of six 10.4 m
telescopes located in the Owens Valley, CA and operated by Caltech}
and BIMA\footnote{An array of ten 6.1 m mm-wave telescopes located at
Hat Creek, California and operated by the
Berkeley-Illinois-Maryland-Association} mm-wave arrays in
California. They have produced \sze\ images toward 60 clusters to date
\citep{carlstrom96, carlstrom99, patel00, grego00a, reese00, grego00b,
joy01, laroque01, reese02}.  A sample of their \sze\ images is shown
in Figure~\ref{fig:szpanel12}. Figure~\ref{fig:szpanel12} also clearly
demonstrates the independence of the \sze\ on redshift. All of the
clusters shown have similar high X-ray luminosities and, as can be
seen, the strength of the \sze\ signals are similar despite the factor
of five in redshift.  The OVRO and BIMA arrays support two dimensional
configuration of the telescopes, including extremely short baselines,
allowing good synthesized beams for imaging the SZE of clusters at
declinations greater than $\sim -15$~degrees.

The Ryle Telescope, OVRO, and BIMA SZE observations are insensitive to
the angular scales required to image low redshift clusters, $z <
<0.1$.  Recently, however, the Cosmic Background Imager (CBI)
\citep{padin01} has been used to image the SZE in a few nearby
clusters \citep{udomprasert00}.  The CBI is composed of thirteen 0.9 m
telescopes mounted on a common platform with baselines spanning 1 m to
6 m. Operating in ten 1 GHz channels spanning 26 - 36 GHz, it is
sensitive to angular scales spanning 3$'$ to 20$'$.  The large field
of view of the CBI, 0.75~ degrees FWHM, makes it susceptible to
correlated contamination from terrestrial sources, i.e., ground
emission. To compensate, they have adopted the same observing strategy
as for single dish observations (\S\ref{sec:single_dish}), and
subtract 
data from leading and trailings fields
offset by $\pm$12.5 minutes in Right Ascension from the cluster.

Interferometric observations of the SZE, as for single dish
observations, are just beginning to demonstrate their
potential. Upcoming instruments will be over an order of magnitude
more sensitive. The OVRO - BIMA SZE imaging team is now building the
Sunyaev-Zel'dovich Array (SZA), consisting of eight 3.5 m telescopes
outfitted with 26 - 36 GHz and 85 - 115 GHz low-noise receivers and
employing an 8 GHz wideband correlator.  The SZA is expected to be
operational by the end of 2003. It will be deployed with the existing
six 10.4 m OVRO telescopes and nine 6.1 BIMA telescopes at a new high
site if the site is ready in time.  If not, the array will be deployed
in the Owens Valley with the existing OVRO telescopes. The SZA will
operate both in a dedicated survey mode and also as a fully
heterogeneous array with the larger telescopes. The heterogeneous
array will provide unprecedented imaging of the SZE at high
resolution.

The Ryle Telescope (RT) SZE team is also building the ArcMinute Imager
(AMI), consisting of ten 3.7m telescopes operating at 15 GHz near the
RT in Cambridge. It is not planned to operate AMI as a heterogeneous
array with the Ryle telescope, but the RT would be used for concurrent
point source monitoring.

Additionally, plans have been discussed to reconfigure the CBI to 90
GHz. With its thirteen 0.9 m telescopes and 10 GHz bandwidth, the CBI
would be a formidable SZE survey machine.  A similar fixed platform
interferometer, the Array for Microwave Background Anisotropy (AMiBA),
is also being built with nineteen 1.2 m telescopes and operating at 90
GHz. AMiBA, like the reconfigured CBI, would also be ideally suited
for performing SZE surveys at moderate resolution.

This next generation of interferometric SZE instruments will conduct
deep SZE surveys covering tens and possibly hundreds of square
degrees. While not as fast as planned large format bolometric arrays,
the interferometers will be able to 
provide more
detailed imaging.  In particular, the high resolution and deep imaging
provided by the SZA/OVRO/BIMA heterogeneous array (referred to as
CARMA, the Combined ARray for Millimeter Astronomy) operating at 90
GHz will provide a valuable tool for investigating cluster structure
and its evolution. As discussed in \S\ref{sec:interpret}, such studies
are necessary before the full potential of large SZE surveys for
cosmology can be realized.

%%%%%%%%%%%%%%%%%%%%%%%%%%%%%%%%%%%%%%%%%%%%%%%%%%%%%%%%%%%%%%%%%%%%%%%%%%%%%
%%%  									  %%%
%%%	IV. SZE Survey Yields						  %%%
%%%  									  %%%
%%%%%%%%%%%%%%%%%%%%%%%%%%%%%%%%%%%%%%%%%%%%%%%%%%%%%%%%%%%%%%%%%%%%%%%%%%%%%

\section{SKY SURVEYS WITH THE SZE}
\label{sec:survey_yields}

With recent developments in instrumentation and observing strategies,
it will soon be possible to image large areas with high sensitivity,
enabling efficient and systematic SZE searches for galaxy clusters.

The primary motivation for large surveys for galaxy clusters using the
SZE is to obtain a cluster catalog with a well understood selection
function that is a very mild function of cosmology and redshift. There
are two primary uses for such a catalog. The first is to use clusters
as tracers of structure formation, allowing a detailed study of the
growth of structure from $z \sim 2$ or 3 to the present day. The
second use is for providing a well-understood sample for studies of
individual galaxy clusters, either as probes of cosmology or for
studies of the physics of galaxy clusters.

Numerous authors have presented estimates of the expected yields from
SZE surveys \citep{korolev86,bond91a,bartlett94,markevitch94,deluca95,
bond96,barbosa96,colafrancesco97,aghanim97,kitayama98,holder00,
bartlett00,kneissl01}. Results from the diverse approaches to
calculating the cluster yields are in broad agreement.

\subsection{Cluster Abundance}
\label{subsec:cluster_abund}

The number of clusters expected to be found in SZE surveys depends
sensitively on the assumed cosmology and detector specifications.
Estimates of the order of magnitude, however, should be robust and
able to give a good indication of the expected scientific yields of
surveys for galaxy clusters using the SZE.

In calculating the number of clusters expected in a given survey,
three things are needed: \newline 1) the volume per unit solid angle
as a function of redshift, \newline 2) the number density of clusters
as a function of mass and redshift, and \newline 3) an understanding of the
expected mass range which should be observable with the particular SZE
instrument and survey strategy.

The physical volume per unit redshift per unit solid angle is given by
\citep{peebles94}
\begin{equation}
{dV \over d\Omega dz} = \Da^{\!\! 2} \ c {dt \over dz} \quad ,
\end{equation} 
where $dt/dz=1/[H(z) (1+z)]$ and $H(z)$ is the expansion rate of the
universe. The comoving volume is simply the physical volume multiplied
by $(1+z)^3$.

The number density of clusters as a function of mass and redshift can
either be derived by applying the statistics of peaks in a Gaussian
random field \citep{press74, bond91, sheth01} to the initial density
perturbations or taken from large cosmological N-body simulations
\citep{jenkins01}.  The mass function is still not understood
perfectly, with small but important differences between competing
estimates, especially at the high mass end of the spectrum. Precise
cosmological studies will require an improved understanding, but
reasonably accurate results can be obtained with the ``standard''
Press-Schechter \citep{press74} mass function, with the comoving
number density between masses $M$ and $M+dM$ given by
\begin{equation}
{dn(M,z) \over dM} = -\sqrt{2 \over \pi} {\bar{\rho} \over M^2} 
{d \ln \sigma(M,z) \over d \ln M} {\delta_c \over \sigma(M,z)} \exp 
\left [ {- \delta_c^2 \over 2 \sigma(M,z)^2} \right ] \quad .
\end{equation} 
In the above, $\bar{\rho}$ is the mean background density of the
universe today, $\sigma^2(M,z)$ is the variance of the density field
when smoothed on a mass scale $M$, and $\delta_c$ (typically $\sim
1.69$) is the critical overdensity for collapse in the spherical
collapse model \citep{peebles80}.

Smoothing the density field on a mass scale corresponds to finding the
comoving volume that encloses a given mass for a region at the mean
density of the universe and smoothing the density field over this
volume.  The variance $\sigma^2(M,z)$ is separable as
$\sigma(M,z)\equiv \sigma(M) D(z)$, where $\sigma(M)$ is the variance
in the initial density field and $D(z)$ is the linear growth function
that indicates how the amplitude of the density field has grown with
time.  For a universe with $\Omega_M=1$ this growth function is simply
proportional to the scale factor.

For a universe composed only of matter and vacuum energy (either with
or without spatial curvature), accurate fitting functions for the
growth function can be found in \citet{carroll92} or can be found as a
straightforward one-dimensional numerical integral, using the solution
found by \citet{heath77}. For a more exotic universe, for example one
with dark energy not in the form of a cosmological constant, the
growth function requires solution of a two dimensional ordinary
differential equation.

The Press-Schechter formulation has the advantage of making it clear
that the abundance of very massive objects is exponentially
suppressed, showing that massive clusters are expected to be rare. The
amount of suppression as a function of redshift is sensitive to the
linear growth function $D(z)$, which is itself sensitive to
cosmological parameters. Structure grows most efficiently when the
universe has $\Omega_M \sim 1$, so the growth function as a function
of $z$ should give a good indication of the epoch when either
curvature, vacuum energy or dark energy started to become
dynamically important. 

\begin{figure}[!tbh]
\centerline{
  \psfig{figure=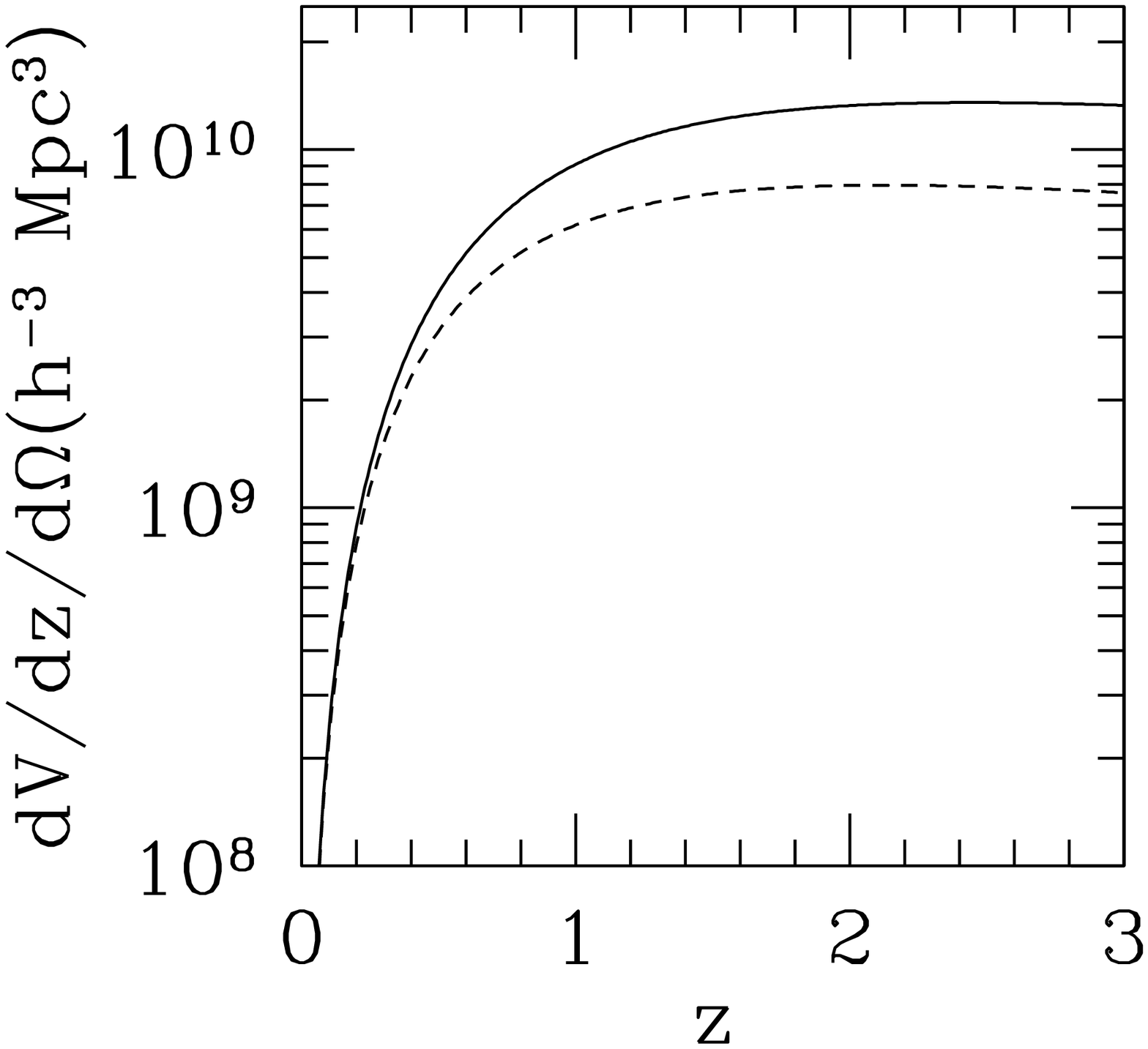,height=1.8in}
  \psfig{figure=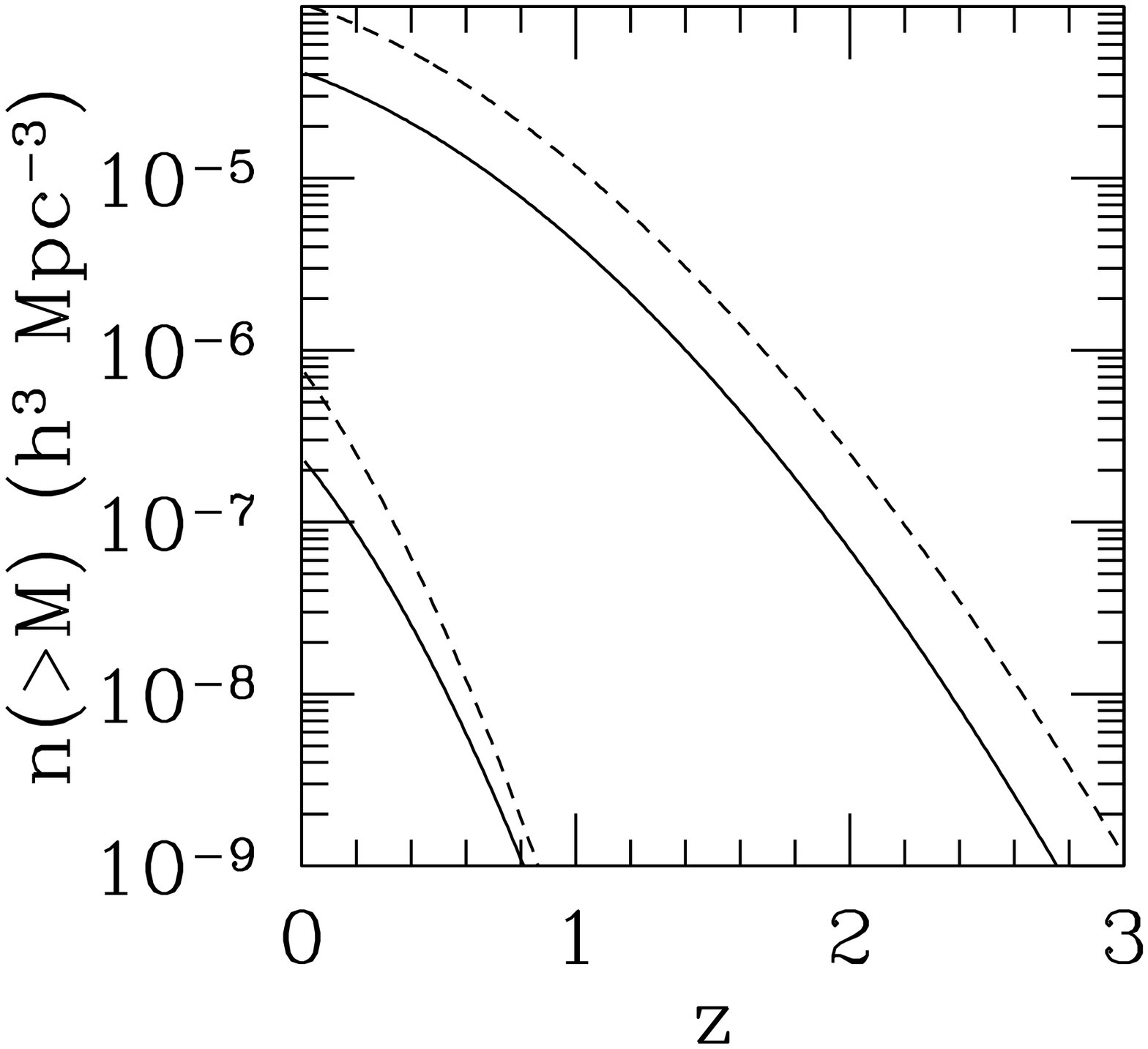,height=1.8in}
  \psfig{figure=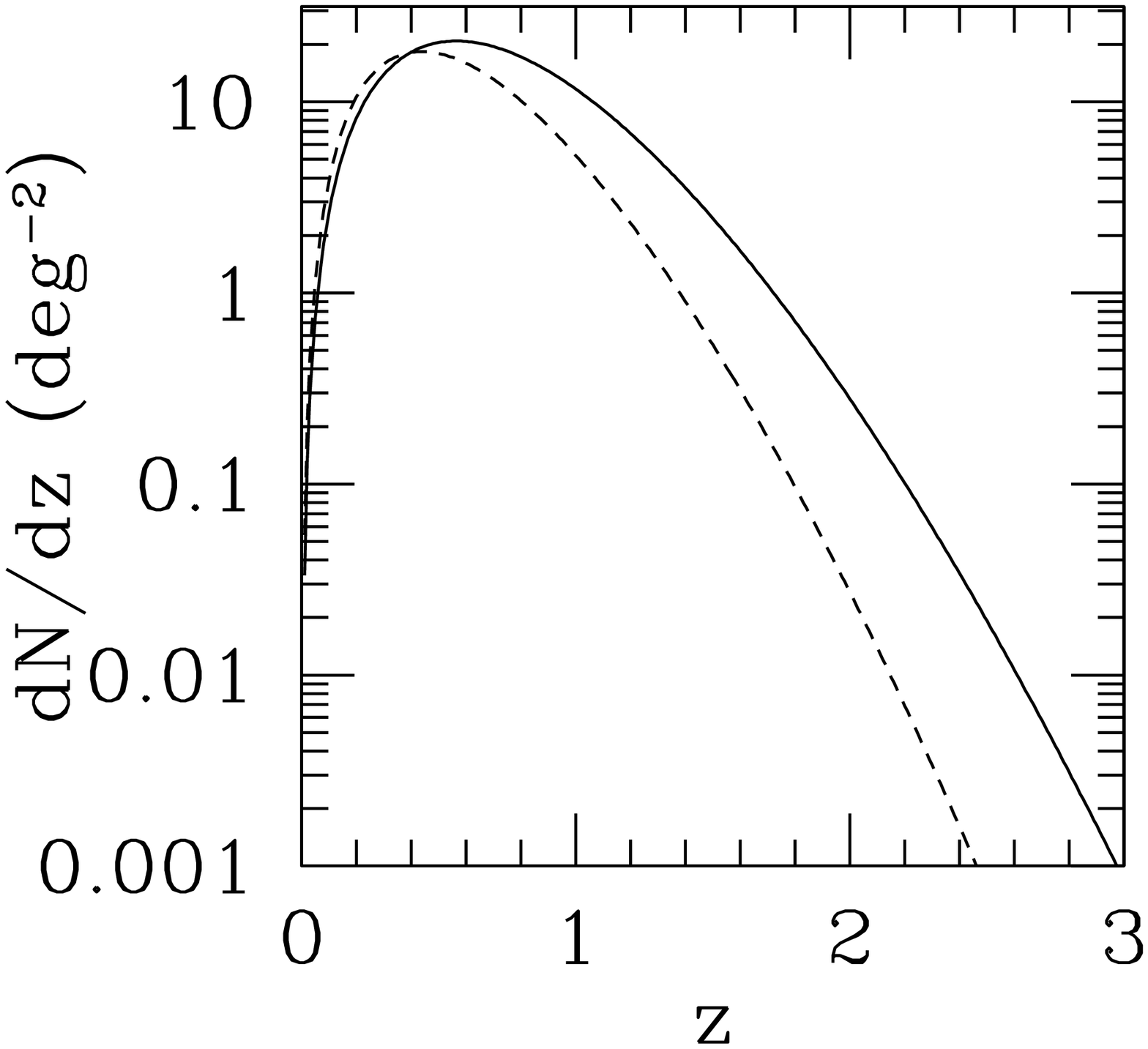,height=1.8in}
}
\caption{Comoving volume element (left) and comoving number
density (center) for two cosmologies, (\Om,\Ol)=(0.3,0.7) (solid)
and (0.5,0.5) (dashed). For the middle panel, the normalization
of the matter power spectrum was taken to be $\sigma_8=0.9$ and
the Press-Schechter mass function was assumed. The lower set of
lines in the middle panel correspond to clusters with mass greater than 
$10^{15} h^{-1}M_\odot$ while the upper lines correspond to clusters
with mass greater than $10^{14} h^{-1}M_\odot$. The right panel corresponds
to the cluster redshift distribution per square degree for clusters with
masses greater than $10^{14} h^{-1}M_\odot$, 
with the normalization of the power spectrum
adjusted ($\sigma_8=0.75$ for $\Om=0.5$)
to produce the same local cluster abundance for
both cosmologies. Note that in this case, fewer clusters
are predicted at high redshift for the higher density
cosmology.
}
\label{fig:vol_vs_n}
\end{figure}

The exponential dependence of the cluster abundance makes SZE surveys
a potentially powerful probe of cosmology.  This is shown in
Figure~\ref{fig:vol_vs_n}, where the relative importance of volume and
number density can be seen. A difference in cosmology can cause a
difference in volume of a few tens of percent, while the corresponding
change in comoving number density is typically a factor of a few. This
plot also shows the rapid decline in the number density of the cluster
abundance with redshift and its steep dependence on mass, both of
which are due to the exponential suppression of high peaks.

The cosmology with the higher mass density can be seen to have a
higher abundance at $z=0$ in Figure~\ref{fig:vol_vs_n} for a fixed
normalization of the power spectrum, i.e., $\sigma_8$.  This is
primarily because a given cluster mass will correspond to a slightly
smaller size for a universe with higher matter density.  The matter
power spectrum rises toward smaller scales, so a fixed amplitude of
the power spectrum on a specific scale will lead to a higher density
cosmology having more power on a given mass scale. Choosing a slightly
lower value of $\sigma_8$ for the cosmology with the higher density
removes this offset in the cluster abundance at $z=0$ and leads to a
lower cluster abundance at higher redshifts, i.e., if the cluster
abundance is normalized at $z = 0$ for all cosmological models, the
higher density models will have relatively fewer clusters at high
redshift.  This can be seen in the right panel of
Figure~\ref{fig:vol_vs_n}, where the redshift distribution per square
degree has been normalized to give the same number of clusters above a
mass of $10^{14} h^{-1}M_\odot$ at $z=0$ by lowering the normalization
of the power spectrum from $\sigma_8=0.90$ to 0.75 for the cosmology with
the higher matter density.

\subsection{Mass Limits of Observability}
\label{subsec:mass_limit}

The range of masses to which a survey is sensitive is set by the
effective beam size and sensitivity of the instrument as well as the
cluster profile on the sky.  In the case of a beam that is larger than
the cluster, a survey is limited by SZE flux. From
equation~\ref{eq:tot_sze_flux}, a flux limit corresponds to constant
$N_e T/\Da^{\!\! 2}$.  The dependence on angular diameter distance
rather than luminosity distance leads to a relative factor of
$(1+z)^4$ when compared to a usual flux limit for emission from a
distant source.  Past $z \sim 1$ the angular diameter distance is
slowly varying, and the gas temperature for a fixed mass should be
higher than at $z=0$, since the clusters are more dense and therefore
more tightly bound, i.e., smaller. As a result, at $z>1$ the limiting
mass for an SZE survey is likely to be gently declining with redshift
\citep{holder00, bartlett00,kneissl01}.  Nearby clusters ($z<0.2$) are
likely to be at least partially resolved by most SZE surveys, making
the mass selection function slightly more difficult to estimate
robustly. It is not expected that the mass threshold of detectability
should change more than a factor of $\sim 2-3$ for clusters with
$z>0.05$, making an SZE selected catalog remarkably uniform in
redshift in terms of its mass selection function.

The expected cluster profiles are not well known, since there are very
few known clusters at $z>0.5$, and these are much more massive than
the typical clusters expected to be found in deep SZE surveys. The
total SZE flux from a cluster should be fairly robust against changes
in cluster profiles due to substructure or merging. Broadly speaking,
the SZE is providing an inventory of hot electrons. The characteristic
temperature in a cluster is set by virial considerations, since the
electrons are mainly heated by shocks due to infall.  Kinetic energy
of infalling gas should be converted into thermal energy in a time
shorter than the Hubble time, suggesting that the thermal energy per
particle must necessarily be on the order of $GM/R$. Therefore, it is
very difficult to substantially alter the expected SZE flux for a
cluster of a given mass.

\begin{figure}[!tbh]
\centerline{\psfig{figure=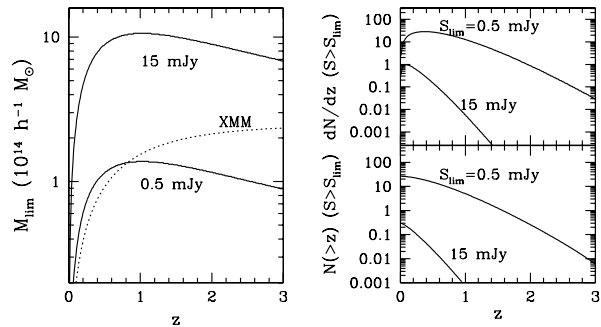,height=2.0in}}
\caption{Left: Mass limits as a function of redshift for a typical
wide-field type of survey (equivalent to $\sim$ 15 mJy at 30 GHz) and
a typical deep survey ($\sim$ 0.5 mJy). The approximate XMM
serendipitous survey limit is also shown. Right: Differential (top)
and cumulative (bottom) counts as a function of redshift for two SZE
surveys shown at left, assuming a $\Lambda$CDM cosmology
\protect\citep{holder00}.
\label{fig:SZE_survey}}
\end{figure}

The mass limit of detection corresponding to an SZE flux limited
survey is shown in the left panel of Figure~\ref{fig:SZE_survey}.  The
two types of surveys shown correspond to deep ground-based imaging of
a few tens of square degrees down to $\mu K$ sensitivities with
arcminute resolution or wide-field surveys (similar to the {\it Planck
Surveyor} satellite) with $\sim 5' - 10'$ resolution.

In contrast to the total integrated SZE flux, the concentration of SZE
flux is very model dependent, with very compact clusters having high
central decrements (or increments) but subtending a relatively small
solid angle.  The integrated SZE flux is thus a potentially very
powerful criterion for controlling selection effects in samples for
cluster studies. The clusters can first be found using integrated SZE
flux and then investigated with high resolution SZE imaging.

\subsection{Estimates of SZE Source Counts}
\label{subsec:sze_counts}

The expected source counts are shown in the right panels of
Figure~\ref{fig:SZE_survey}. From the considerations discussed above,
it should be clear that the exact numbers will depend on cosmology and
observing strategy.  A robust conclusion is that upcoming deep surveys
should find tens of clusters per square degree. Less deep surveys,
such as the all-sky {\it Planck Surveyor} satellite survey should
detect a cluster in every few square degrees.  The resulting catalogs
should be nearly uniformly selected in mass, with the deep catalogs
extending past $z\sim2$.

%%%%%%%%%%%%%%%%%%%%%%%%%%%%%%%%%%%%%%%%%%%%%%%%%%%%%%%%%%%%%%%%%%%%%%%%%%%%%
%%%  									  %%%
%%%	V. Cosmology from SZE Survey Samples				  %%%
%%%  									  %%%
%%%%%%%%%%%%%%%%%%%%%%%%%%%%%%%%%%%%%%%%%%%%%%%%%%%%%%%%%%%%%%%%%%%%%%%%%%%%%

\section{COSMOLOGY FROM SZE SURVEY SAMPLES}
\label{sec:sze_cosmo}

In this section we review the use of the SZE for cosmological studies
and provide an update on current constraints. Emphasis is given to the
cosmology that can, in principle, be extracted from SZE survey yields,
as well as the observational and theoretical challenges which must be
met before the full potential of SZE surveys for cosmology can be
realized.

Future SZE surveys, with selection functions that are essentially a
uniform mass limit with redshift (\S\ref{sec:survey_yields}), will
provide ideal cluster samples for pursuing cosmology. A large catalog
of distant clusters will enable studies of large scale structure using
the same methods as are applied to large catalogs of galaxies. The SZE
surveys will also provide a direct view of the high redshift universe.
Should clusters exist at redshifts much higher than currently
predicted, i.e., if the initial mass fluctuations were non-Gaussian,
they will be found by SZE surveys, but missed in even the deepest
X-ray observations planned.

The SZE survey samples can be used to increase the precision of the
more traditional applications of the SZE to extract cosmological
parameters, such as cluster distance measurements and the Hubble
constant, the ratio \Ob/\Om, and cluster peculiar velocities.  These
are discussed in \S\ref{sec:cosmo_hubble}, \S\ref{sec:cosmo_f_gas} and
\S\ref{sec:cosmo_vpec}, respectively.  The ability to derive these
parameter depends primarily on the ability to use the SZE and other
cluster observables to constrain or even overconstrain cluster
properties. Precise SZE measurements will allow tests of the
underlying assumptions in these derivations.  For example, high
resolution imaging of SZE, X-ray, and lensing, will allow detailed
tests of the assumption of hydrostatic equilibrium
\citep{miralda-escude95b, loeb94, wu97, squires96, allen98}.  The
sample yields will also allow the determination of global properties
of clusters and their relationship to observables.  For example, it is
already possible to estimate cluster gas temperatures without X-ray
data using current SZE data \citep{joy01}.

The new frontier for SZE cosmology will be in exploiting the ability
of future SZE surveys to measure cleanly the number density of
clusters and its evolution in time.  The redshift distribution of
galaxy clusters is critically sensitive to \Om\ and the properties of
the dark energy. For sufficiently large and deep SZE surveys, it is
possible, in principle, to extract the equation of state of the dark
energy.  This is discussed in \S\ref{sec:survey_science} and the
theoretical and observational challenges are outlined in
\S\ref{sec:interpret}.

\subsection{Distance Determinations, Hubble Constant}
\label{sec:cosmo_hubble}

Several years after the SZE was first proposed \citep{sunyaev70,
sunyaev72} it was recognized that the distance to a cluster could be
determined with a measure of its SZE and X-ray emission
\citep{cavaliere77, gunn78, silk78, cavaliere78,
birkinshaw79}.  The distance is determined by exploiting the different
density dependencies of the SZE and X-ray emission.  The SZE is
proportional to the first power of the density; $\Delta T_{SZE} \sim
\int d\ell n_e T_e$, where $n_e$ is the electron density, $T_e$ is the
electron temperature, and $d\ell$ is along the line-of-sight.  The
distance dependence is made explicit with the substitution $d\ell =
\Da d\zeta$, where $\Da$ is the angular diameter distance of the
cluster.

\begin{figure}[!tbh]
   \centerline{\psfig{figure=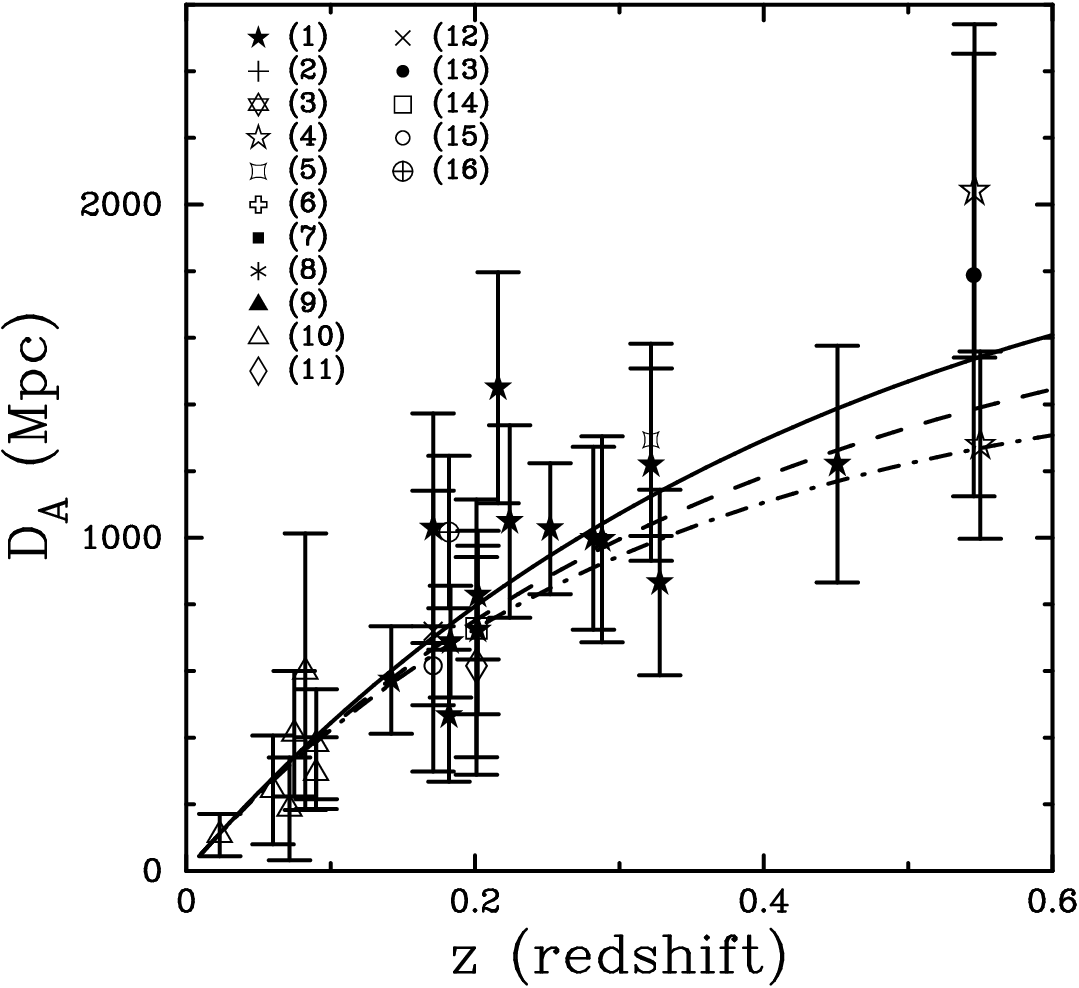,height=3in}}
   \caption{SZE determined distances versus
   redshift. The theoretical angular diameter distance relation is
   plotted for three different cosmologies, assuming $\Ho = 60$ \ksM:
   $\Lambda$ $\OmM=0.3$, $\OmL=0.7$ (solid line); open $\OmM=0.3$
   (dashed); and flat $\OmM=1$ (dot-dashed).
   The clusters are beginning to trace out the angular diameter
   distance relation.  
References: 
	(1)  \protect\citealt{reese02};
	(2)  \protect\citealt{pointecouteau01};
	(3)  \protect\citealt{mauskopf00};
	(4)  \protect\citealt{reese00};
	(5)  \protect\citealt{patel00};
	(6)  \protect\citealt{grainge00};
	(7)  \protect\citealt{saunders00};
	(8)  \protect\citealt{andreani99};
	(9)  \protect\citealt{komatsu99};
	(10) \protect\citealt{mason01, mason99, myers97};
	(11) \protect\citealt{lamarre98};
	(12) \protect\citealt{tsuboi98};
	(13) \protect\citealt{hughes98b};
	(14) \protect\citealt{holzapfel97};
	(15) \protect\citealt{birkinshaw94};
	(16) \protect\citealt{birkinshaw91b}.
\label{fig:da}}
\end{figure}

The X-ray emission is proportional to the second power of the density;
$\Sx \sim \int d\ell n_e^2 \LameH$, where \LameH\ is the X-ray cooling
function.  The angular diameter distance is solved for by eliminating
the electron density\footnote{Similarly, one could eliminate \Da\ in
favor of the central density, \neo} yielding
\begin{equation}
\Da \propto \frac{(\dTo)^2 \Lamo}{\Sxo T_{e 0}^2} \frac{1}{\theta_c},
	\label{eq:Dadepend}
\end{equation}
where these quantities have been evaluated along the line of sight
through the center of the cluster (subscript 0) and $\theta_c$ refers
to a characteristic scale of the cluster along the line of sight,
whose exact meaning depends on the density model adopted.  Only the
characteristic scale of the cluster in the plane of the sky is
measured, so one must relate the line of sight and plane of sky
characteristic scales.  For detailed treatments of this calculation
see \citet{birkinshaw94} and \citet{reese00}.  Combined with the
redshift of the cluster and the geometry of the universe, one may
determine the Hubble parameter, with the inverse dependencies on the
observables as that of $\Da$.  With a sample of galaxy clusters, one
fits the cluster distances versus redshift to the theoretical angular
diameter distance relation, with the Hubble constant as the
normalization, e.g., see Figure~\ref{fig:da}.
  
There are two explicit assumptions made in SZE and X-ray distance
determinations. The first one, mentioned above, is that the
characteristic scale of the cluster along the line of sight must be
related (usually assumed equal) to the scale in the plane of the sky.
Typically, spherical symmetry is assumed for the cluster geometry
since for a large sample of clusters one would expect
$\left<\theta_c/\theta_c^{sky}\right> = 1$, at least in the absence of
selection effects.  This assumption is supported by simulations as
well \citep{sulkanen99}. The second assumption is that $\left<n_e^2
\right >^{1/2}$ equals $\left < n_e \right >$ along the line of sight,
i.e., that the clumping factor
\begin{equation}
C \equiv \frac{\left < n_e^2 \right >^{1/2}}{\left < n_e \right >},
\label{eq:C_def}
\end{equation}
equals unity.  If significant substructure exists in galaxy clusters,
the derived Hubble constant will be overestimated by a factor of
$C^2$.

To date, there are 38 distance determinations to 26 different galaxy
clusters from analyses of \sze\ and X-ray observations.  In
Figure~\ref{fig:da} we show all SZE determined distances from high
signal-to-noise SZE experiments.  The uncertainties shown are
statistical at 68\% confidence.  There are currently three samples of
clusters with SZE distances: 1) a sample of 7 nearby ($z<0.1$) galaxy
clusters observed with the OVRO 5m telescope \citep{myers97, mason01};
2) a sample of 5 intermediate redshift ($0.14<z<0.3$) clusters from
the Ryle telescope interferometer \citep{jones01}; and 3) a sample of
18 clusters with $0.14 < z < 0.83$ from interferometric observations
by the OVRO and BIMA SZE imaging project \citep{reese02}.  A fit to the
ensemble of 38 SZE determined distances yields $\Ho = 60 \pm 3$ km
s$^{-1}$ Mpc$^{-1}$ for an $\Om = 0.3$, $\Ol = 0.7$ cosmology, where
only the statistical uncertainty is included (at 68\% confidence).
The systematic uncertainty, discussed below, is of order 30\% and
clearly dominates. Since many of the clusters are at high redshift,
the best fit Hubble constant will depend on the cosmology adopted; the
best fit Hubble constant shifts to 56 km s$^{-1}$ Mpc$^{-1}$ for an
open $\Om = 0.3$ universe and to 54 km s$^{-1}$ Mpc$^{-1}$ for a flat
$\Om = 1$ geometry.

The prospects for improving both the statistical and systematic
uncertainties in the SZE distances in the near future are promising.
Note, from Eq.~\ref{eq:Dadepend}, that the error budget in the
distance determination is sensitive to the absolute calibration of the
X-ray and SZE observations.  Currently, the best absolute calibration
of SZE observations is $\sim 2.5$\% at 68\% confidence based on
observations of the brightness of the planets Mars and Jupiter.
Efforts are now underway to reduce this uncertainty to the 1\% level
(2\% in \Ho).  Uncertainty in the X-ray intensity scale also adds
another shared systematic. The accuracy of the ROSAT X-ray intensity
scale is debated, but a reasonable estimate is believed to be $\sim
10$\%. It is hoped that the calibration of the Chandra and XMM-Newton
X-ray telescopes will greatly reduce this uncertainty.

The largest systematic uncertainties are due to departures from
isothermality, the possibility of clumping, and possible point source
contamination of the \sze\ observations \citep[for detailed discussion
of systematics see, e.g.,][]{birkinshaw99, reese00, reese02}.  Chandra
and XMM-Newton are already providing temperature profiles of galaxy
clusters \citep[e.g.,][]{nevalainen01, markevitch01, tamura01}.  The
unprecedented angular resolution of Chandra will provide insight into
possible small scale structures in clusters.  In addition,
multi-wavelength studies by existing radio observatories, e.g., the
Very Large Array (VLA), can shed light on the residual point source
contamination of the radio wavelength SZE measurements.  Therefore,
though currently at the 30\% level, many of the systematics can and
will be addressed through both existing X-ray and radio observatories
and larger samples of galaxy clusters provided from SZE surveys.

The beauty of the SZE and X-ray technique for measuring distances is
that it is completely independent of other techniques, and that it can
be used to measure distances at high redshifts directly.  Since the
method relies on the well understood physics of fully ionized plasmas,
it should be largely independent of cluster evolution.  Inspection of
Figure~\ref{fig:da} already provides confidence that a large survey of
SZE distances consisting of perhaps a few hundred clusters with
redshifts extending to one and beyond would allow the technique to be
used to trace the expansion history of the universe, providing a
valuable independent check of the recent determinations of the
geometry of the universe from type Ia supernova \citep{riess98,
perlmutter99} and CMB primary anisotropy experiments
\citep{pryke01,netterfield01, stompor01}.

\subsection{Cluster gas mass fractions,  $\Omega_M$ }
\label{sec:cosmo_f_gas}

The ICM contains most of the baryons confined to the cluster potential
with roughly an order of magnitude more baryonic mass than that
observed in the galaxies themselves \citep{white93, forman82}. The gas
mass fraction, $f_g$, is therefore a reasonable estimate of the
baryonic mass fraction of the cluster. It should also be reasonable
approximation to the universal baryon mass fraction, $\fb \equiv
\OmB/\OmM$, since it is not believed that mass segregation occurs on
the large scales from which massive clusters condense $\sim 1000$
Mpc$^3$.  The cluster gas fraction is actually a lower limit, $f_g \le
\fb$, since a small fraction of baryons ($\sim 10$\%) are likely lost
during the cluster formation process \citep{white93,evrard97}, and we
can not rule out the possibility of additional reservoirs of baryons
in galaxy clusters which have yet to be detected.

A measurement of $\fb$ leads directly to an estimate of $\OmM$ given a
determination of $\OmB$.  Recent reanalysis of big bang
nucleosynthesis predictions with careful uncertainty propagation
\citep{burles01,nollett00, burles99} along with recent D/H measurements in
Lyman $\alpha$ clouds \citep{burles98a, burles98b} constrain the
baryon density to be $\Omega_B h^2 = 0.020 \pm 0.002$ at 95\%
confidence.  Recent CMB primary anisotropy experiments provide an
additional independent determination of $\Omega_B h^2$ consistent with
the Lyman $\alpha$ cloud result
\citep{pryke01,netterfield01,stompor01}.

The gas mass is measured directly by observations of the SZE provided
the electron temperature is known.  The total gravitating mass can be
determined by assuming hydrostatic equilibrium and using the
distribution of the gas and, again, the electron temperature.  The SZE
derived gas fraction will therefore be proportional to $\Delta T_{SZE}
/ T_e^2$. Alternatively, the total gravitating mass can be determined
by strong lensing (on small scales) or weak lensing (on large scales).
Recently there has been considerable work on SZE gas fractions using
total mass determinations derived under the assumption of hydrostatic
equilibrium.

\sze\ derived cluster gas mass fractions have been determined for two
samples of clusters and the results were used to place constraints on
$\Omega_M$: a sample of four nearby clusters \citep{myers97} and a
sample of 18 distant clusters \citep{grego00b}. Both analyses used a
spherical isothermal $\beta$-model for the ICM.  The nearby sample was
observed with the Owens Valley 5.5 m telescope at 32 GHz as part of a
SZE study of an X-ray flux limited sample \citep{myers97}. In this
study, the integrated SZE was used to normalize a model for the gas
density from published X-ray analyses, and then compared to the
published total masses to determine the gas mass fraction.  For three
nearby clusters, A2142, A2256 and the Coma cluster, a gas mass
fraction of $f_g h = 0.061 \pm 0.011$ at radii of 1-1.5 $h^{-1}$ Mpc
is found; for the cluster Abell 478, a gas mass fraction of $f_g h =
0.16 \pm 0.014$ is reported.

The high redshift sample of 18 clusters ($0.14 < z < 0.83$) was
observed interferometrically at 30 GHz using the OVRO and BIMA SZE
imaging system \citep{grego00b}. In this study, the model for the gas
density was determined directly by the SZE data.  X-ray emission
weighted electron temperatures were used, but no X-ray imaging data
was used.  The gas fractions were computed from the data at a
1$^\prime$ radius where they are best constrained by the observations.
Numerical simulations suggest, however, that the gas mass fraction at
$r_{500}$ (the radius inside of which the mean density of the cluster
is 500 times the critical density) should reflect the universal baryon
fraction \citep{evrard97, evrard96, david95}.  The derived gas
fractions were therefore extrapolated to $r_{500}$ using scaling
relations from cluster simulations \citep{evrard97}.  The resulting
mean gas mass fractions are $f_g h = 0.081 ^{+0.009} _{-0.011}$ for
$\OmM = 0.3,\ \OmL=0.7$, $f_g h = 0.074^{+0.008} _{-0.009}$ for $\OmM
= 0.3,\ \OmL=0.0$ and $f_g h = 0.068 ^{+0.009} _{-0.008}$ for $\OmM =
1.0,\ \OmL=0.0$.  The uncertainties in the electron temperatures
contribute the largest component to the error budget.

\begin{figure}[!tbh]
\centerline{\psfig{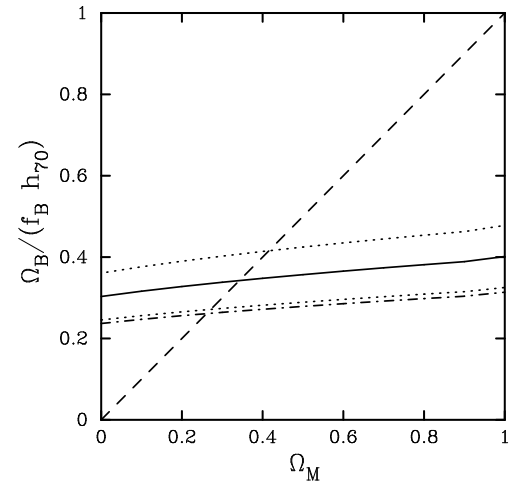}}
\caption{Limits on $\Omega_M$ from SZE measured cluster gas
fractions \protect\citep{grego00b}. Upper limit on the total matter density,
$\Omega_M \le \Omega_B/(\fb h_{70})$ (solid line) and its associated
68\% confidence region (dotted lines), as a function of cosmology with
$\Omega_\Lambda \equiv 1 - \Omega_M$.  The intercept between the upper
dotted line and the dashed line $\Omega_M = \Omega_B/(\fb h_{70})$
gives the upper limit to $\Omega_M$ at 68\% confidence.  The
dot-dashed line shows the total matter density when the baryon
fraction includes an estimate of the contribution from baryons in
galaxies and those lost during cluster formation.  The intercept of
the dot-dashed line and the dashed line gives the best estimate of
$\Omega_M \sim 0.25$ assuming a flat universe with $h =0.7$.}
\label{fig:fb_omega}
\end{figure}

The angular diameter distance relation, $\Da(z)$, enters the gas
fraction calculation and introduces a cosmology dependence on the
results of the high $z$ sample.  In addition, the simulation scaling
relations used to extrapolate the gas fractions to $r_{500}$ have a
mild dependence on cosmology.  Figure~\ref{fig:fb_omega} shows the
constraints on $\Omega_M$ implied by the measured gas mass fractions
assuming a flat universe ($\Omega_\Lambda \equiv 1- \Omega_M$) and
$h=0.7$ to calculate $\Da$ and the $r_{500}$ scaling factor.  The
upper limit to $\Omega_M$ and its associated 68\% confidence interval
is shown as a function of $\Omega_M$.  The measured gas mass fractions
are consistent with a flat universe and $h=0.7$ when $\Omega_M$ is
less than 0.40, at 68\% confidence.  For the measurements to be
consistent with $\Omega_M = 1.0$ in a flat universe, the Hubble
constant must be very low, $h$ less than $\sim 0.30$.

To estimate \OmM, we need to account for the baryons contained in the
galaxies and those lost during cluster formation. The galaxy
contribution is assumed to be a fixed fraction of the cluster gas,
with the fraction fixed at the value observed in the Coma cluster,
$M_{g}^{true} = M_g^{obs}(1+0.20 h^{3/2})$
\citep{white93}. Simulations suggest that the baryon fraction at
$r_{500}$ will be a modest underestimate of the true baryon fraction
$f_g(r_{500}) = 0.9\times \fb(\mbox{universal})$
\citep{evrard97}. These assumptions lead to $\fb = [f_g (1 +
0.2h^{3/2})/ 0.9]$.  Using this to scale the gas fractions derived
from the high $z$ SZE cluster sample and assuming $h=0.7$ and a flat
cosmology, leads to the constraints illustrated in
Figure~\ref{fig:fb_omega} with a best estimate $\OmM \sim 0.25$
\citep{grego00b}.

Cluster gas mass fractions can also be determined from cluster X-ray
emission in a similar manner as from SZE measurements.  However, there
are important differences between X-ray and SZE determined gas
fractions. For example, the X-ray emission is more susceptible to
clumping of the gas, $C$, since it is proportional to the ICM density
squared. On the other hand, the X-ray derived gas mass is essentially
independent of temperature for the ROSAT 0.1-2.4 keV band used in the
analyses \citep{mohr99}, while the SZE derived gas mass is
proportional to $T_e^{-2}$.

Currently X-ray data for low redshift clusters is of exceptional
quality, far surpassing SZE data.  X-ray based gas mass fractions have
been measured to cluster radii of 1 Mpc or more
\citep[e.g.,][]{white95, david95, neumann97, squires97, mohr99}.  A
mean gas mass fraction within $r_{500}$ of $f_g h^{3/2} = 0.0749 \pm
0.0021$ at 90\% confidence was derived from X-ray data from a large,
homogeneous, nearby sample of clusters with $T_e > 5$ keV
\citep{mohr99}.  The gas mass fractions derived from SZE measurements
depend differently on the cosmology assumed than those derived from
X-ray images, and this should be noted when comparing the results.
Qualitatively, the comparison does not suggest any large systematic
offsets. In fact, for a $\Lambda$CDM cosmology, solving for $h$ from
the combination of the \citet{grego00b} and \citet{mohr99} results
gives $h = 0.85^{+0.30}_{-0.20}$ at 68\% confidence.  This is
significant, because a large clumping factor, $C \gg 1$ (see
Eq.~\ref{eq:C_def}), has been suggested as an explanation for the high
gas mass fractions in clusters \citep{white93, evrard97}.

Cluster gas mass fractions can also be measured by comparing \sze\
derived gas masses and weak lensing derived total masses.  The
comparison is particularly interesting as both are measures of
projected mass distributions. In addition, gas mass fractions can be
derived without assuming a model for the cluster structure and without
assuming hydrostatic equilibrium. Comparisons of \sze\ and lensing
data has only been done for a few clusters to date \citep{holder00c}.
However, as for the \sze, the quality and quantity of weak lensing
observations toward galaxy clusters is rapidly increasing and several
weak lensing surveys are underway.  \citet{holder00c} demonstrated
that gas mass fractions can be determined from the analysis of \sze\
and weak lensing measurements without need to parameterize the ICM
distribution.  Furthermore, by comparing this mass fraction with one
derived by assuming hydrostatic equilibrium, it is possible to solve
for the ICM electron temperature and the angular diameter distance.

SZE surveys will provide a large catalog of galaxy clusters at
redshifts $z > 1$.  The increased sensitivity and larger angular
dynamic range of the next generation of SZE instruments will allow
measurements of cluster gas fractions to $r_{500}$ directly, greatly
increasing the precision of the gas mass fractions.  Moreover,
extending the gas fraction analyses to high redshift will enable
studies of the evolution of cluster structure.  It should, for
example, be straight forward to test speculative theories of dark
matter decay \citep{cen01}.

\subsection{Cluster Peculiar Velocities}
\label{sec:cosmo_vpec}

The kinetic \sze\ is a unique and potentially powerful cosmological
tool as it provides the only known way to measure large scale velocity
fields at high redshift (\S\ref{sec:kinetic_sze}).  To obtain an
accurate measure of the peculiar velocity of a galaxy cluster,
sensitive multifrequency \sze\ observations are required to separate
the thermal and kinetic effects.  From inspection of
Figure~\ref{fig:spectrum}, it is clear that measurements of the
kinetic \sze\ are best done at frequencies near the null of the
thermal effect at $\sim 218$~GHz.  However, as discussed in
\S\ref{sec:confusion}, contamination by CMB temperature fluctuations
as well as other sources make it difficult to determine accurately the
peculiar velocity for a given cluster.  There have been only a few
recent attempts to measure the kinetic \sze.

The first interesting limits on the peculiar velocity of a galaxy
cluster were reported in \citet{holzapfel97b}.  They used the
Sunyaev-Zel'dovich Infrared Experiment (SuZIE) to observe Abell~2163
($z = 0.202$) and Abell~1689 ($z = 0.183$) at 140~GHz (2.1 mm),
218~GHz (1.4 mm) and 270~GHz (1.1 mm). These observations include and
bracket the null in the thermal SZE spectrum.  Using a $\beta$ model,
with the shape parameters ($\theta_c$, $\beta$) from X-ray data, they
found $v_{pec} = +490 ^{+1370} _{-880}$ \kms\ for Abell 2163 and
$v_{pec} = +170 ^{+815}_{-630}$ \kms\ for Abell 1689, where the
uncertainties are at 68\% confidence and include both statistical and
systematic uncertainties.  These results are limited by the
sensitivity of the SZE observations, which were limited by
differential atmospheric emission. The SuZIE data for Abell~2163 were
reanalyzed with the addition of higher frequency measurements which
are sensitive to emission from Galactic dust in the direction of the
cluster \citep{lamarre98}. More recently \citet{laroque01} also
reanalyzed all of the available data for Abell~2163, including a new
measurement obtained with the OVRO and BIMA SZE imaging system at 30
GHz (1 cm). As shown in Figure~\ref{fig:a2163_spec}, the data is well
fitted by parameters similar to the original values from
\citet{holzapfel97b}. The agreement between the measurements using
different instruments and techniques is striking.

The intrinsic weakness of the kinetic \sze\ and the degeneracy of its
spectrum with that of primary CMB fluctuations make it exceeding
difficult to use it to measure the peculiar velocity of a single
cluster. It may be possible, however, to determine mean peculiar
velocities on extremely large scales by averaging over many clusters.

\subsection{Energy Densities in the Universe and Growth of Structure}
\label{sec:survey_science}

The evolution of the abundance of galaxy clusters is a sensitive probe
of cosmology (\S\ref{sec:survey_yields}). Measurements of the clusters
masses and number density as a function of redshift can be used to
constrain the matter density, $\Omega_M$, and, for sufficiently large
samples, the equation of state of the dark energy. X-ray surveys have
already been used to constrain $\Omega_M$ (e.g.,
\citealt{borgani01,viana99, bahcall98, oukbir97}), but they have been
limited by sample size and their reduced sensitivity to high redshift
clusters.  SZE surveys offer the attractive feature of probing the
cluster abundance at high redshift as easily as the local universe; as
discussed in \S\ref{subsec:mass_limit}, the sensitivity of a SZE
survey is essentially a mass limit \citep{bartlett94, barbosa96,
holder00,bartlett00,kneissl01}.

\begin{figure}[!tbh]
\centerline{\psfig{figure=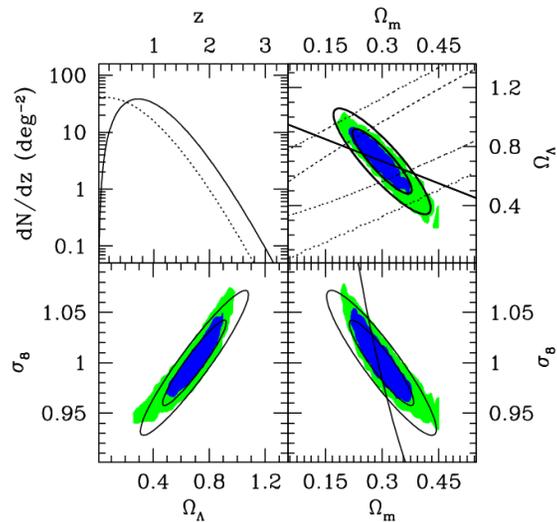,width=0.7\columnwidth}}
\caption{Expected constraints in the $\Omega_M$--$\Omega_\Lambda$ (top
right), $\Omega_M$--$\sigma_8$ (bottom right), and
$\Omega_\Lambda$--$\sigma_8$ (bottom left) planes from the analysis of
a SZE survey covering 12 square degrees in which all clusters above
$10^{14} h^{-1}$ M$_\odot$ are detected and the redshifts are known.
The top left panel shows the expected redshift distribution of
clusters (solid line) with the cumulative $N(>z)$ shown as a dotted
line. In the other panels, the 68\% confidence regions are shown by
the darkest shaded regions, and the 95\% confidence by the lighter
regions.  The solid contours correspond to the same confidence regions
derived from an approximate method.  In each panel the dimension not
shown has been marginalized over rather than kept fixed.  In the upper
right panel, the broken line diagonal ellipses are the constraints
based on the analyses of type Ia Supernova at 68\% and 95\% confidence
\protect\citep{riess98, perlmutter99}. The diagonal line at $\Omega_M +
\Omega_\Lambda = 1$ is for a flat universe as suggested by recent CMB
anisotropy measurements \protect\citep{miller99, debernardis00, hanany00,
pryke01}.  The solid line in lower right panel shows the approximate
direction of current constraints  \protect\citep{viana99}.}
\label{fig:om_ol}
\end{figure}

The simple mass selection function of SZE surveys could allow the
source count redshift distribution to be used as a powerful measure of
cosmological parameters
\citep{barbosa96,haiman00,holder01a,weller01,benson01} and the
structure formation paradigm in general. As an example, we show in
Figure~\ref{fig:om_ol} the expected constraints in
$\Omega_M$--$\Omega_\Lambda$--$\sigma_8$ parameter space from the
analysis of a deep SZE survey \citep{carlstrom99, holder00,holder01a}
covering 12 square degrees in which all clusters above $10^{14}
h^{-1}$ M$_\odot$ are detected and for which redshifts have been
obtained. The darkest region corresponds to the 68\% confidence
region, red to 95\%.  The shaded regions show the result of a Monte
Carlo method for estimating confidence regions. Many realizations of a
fiducial model (\Om=0.3, \Ol=0.7, $\sigma_8$=1) were generated and fit
in the three-dimensional parameter space, and the shaded regions
indicate regions which contain 68\% and 95\% of the resulting best
fits.  The contours show confidence regions from a Fisher matrix
analysis, where the confidence regions are assumed to be Gaussian
ellipsoids in the parameter space. In each case, the
dimension not shown has been marginalized over, rather than kept
fixed.

\begin{figure}[!tbh]
\centerline{\psfig{figure=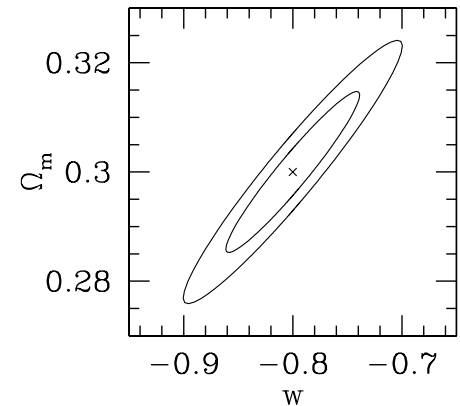,height=3in}}
\caption{Expected constraints on the matter density $\Omega_M$ and the
dark energy equation of state $w$ from the analysis of a SZE survey
covering several thousand square degrees in which all clusters above
$2.5\times 10^{14} h^{-1}$ M$_\odot$ are detected and the redshifts
are known. The normalization of the power spectrum has been
marginalized over, and contours show $68\%$ and $95\%$ confidence
regions for two parameters.  Note, that no systematic errors have been
assumed in deriving the cosmological constraints. As discussed in the
text, considerable observational and theoretical work needs to be done
before such tight constraints could be extracted from large scale
\sze\ surveys. }
\label{fig:sp_om_w}
\end{figure}

The next generation of dedicated telescopes equipped with large format
bolometer detector arrays offers the possibility of conducting \sze\
surveys over thousands of square degrees with $\lsim 10 \mu K$
sensitivity. As shown in Figure~\ref{fig:sp_om_w}, the yields from
such a survey should, in principle (see \S\ref{sec:interpret}), enable
highly accurate estimation of cosmological parameters, notably of the
matter density and the properties of the dark energy.  Most
importantly, the degeneracies in the constraints of the cosmological
parameters derived from SZE survey yields are very different from
those expected from distance measures or CMB measurements. This simply
arises because clusters are probing a fundamentally different physical
effect, the growth rate of structure, rather than distance. Both
growth and distance are related to the expansion history of the
universe, but the two measures are effectively sensitive to different
moments of the expansion rate.

A generic prediction of inflation is that the primordial fluctuations
should be Gaussian. With cluster surveys probing the highest peaks of
the density field, non-Gaussianity in the form of an excess of high
peaks should be readily apparent, especially at high redshift
\citep{benson01}.  Cluster surveys are therefore probing both the
structure formation history of the universe and the nature of the
primordial fluctuations. In this way, cluster surveys are emerging as
the next serious test of the cold dark matter paradigm.

\subsection{Challenges for Interpreting SZE Surveys}
\label{sec:interpret}

In order to realize the full potential of the evolution of the cluster
number density as a cosmological probe, a strong understanding of the
physics of galaxy clusters will be required. As fully collapsed
objects, the complete physics of galaxy clusters is highly non-linear
and complex.  The size of such massive objects, however, makes them
insensitive to disruption from most physical mechanisms.
Nevertheless, there are several important aspects of gas dynamics that
could affect interpretation of SZE galaxy cluster surveys.

\begin{figure}[!tbh]
\centerline{\psfig{figure=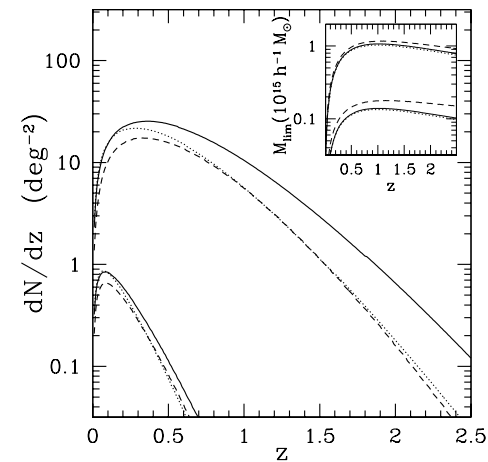,height=3.0in}}
\caption{Effects of gas evolution on cluster survey yields. In the
inset, the top group of lines correspond to mass limits for a SZE
survey similar to the {\it Planck Surveyor} satellite survey, with the
uppermost line indicating the expected mass limit for a model with
significant gas heating while the lower line in the top group shows
the expected mass limit of detection for the case of no cluster gas
heating. The cosmology chosen is \Om=0.3,\Ol=0.7,$\sigma_8=1$. The
lower set of lines show the same effects for a deep SZE survey. The
main panel shows the expected redshift distributions for the mass
limits in the inset.  In the main panel the top group of lines
correspond to the deep surveys while the lower group correspond to
{\it Planck Surveyor}. The solid line within each set shows the
expected counts for the case of no heating, while the dashed line
shows the effect of gas heating.  The dotted curve, i.e., the middle
curve in each set, shows a model with no heating but with \Om=0.33 and
$\sigma_8$ modified to keep the same number of clusters at $z=0$
 \protect\citep{holder01}. The assumptions of no heating and the very high
value of heating for this plot are extreme and should bracket the true
gas evolution.  }
\label{fig:preheat}
\end{figure}

The SZE is only sensitive to free electrons; any process that removes
electrons from the optically thin ICM can affect the magnitude of the
SZE for a given total mass. For example, cooling of the ICM, star
formation or heating of the ICM from supernovae can affect the
observed SZE \citep{holder99a,springel00,holder01,benson01}.  If the
cooling or star formation is not dependent on cluster mass or
redshift, this can be simply calibrated and accounted for in deriving
the survey selection function. The most promising theoretical path for
understanding such processes is through high resolution cosmological
simulations that include the relevant gas dynamics; such simulations
are only now becoming feasible.  The possible effects of such gas
dynamics are shown in Figure~\ref{fig:preheat} \citep{holder01}.  A
simple model has been adopted for this figure, where some combination
of heating or cooling has reduced the number of hot electrons in the
central regions of galaxy clusters by an amount that is modeled
through the effects of an ``entropy floor'' \citep{ponman99}. The
curves in this figure show the extreme cases of either no heating or
cooling (no entropy floor) or extreme gas evolution, with an assumed
value for the minimum entropy that is roughly a factor of two larger
than is required for consistency with the observations
\citep{ponman99}.  Changing the evolution of the ICM could mimic
changes in cosmological parameters at levels much larger than the
expected statistical errors.  With detailed imaging of the ICM using
SZE and X-ray telescopes, the effects of heating or cooling should be
apparent, so the possible systematic errors due to heating or cooling
should in practice be much smaller than $10\%$ in \Om.

Although the SZE-mass relation is easy to understand theoretically in
general terms, the details of the normalization and redshift evolution
will require additional studies of at least a moderately sized SZE
cluster catalog. For example, simple scaling arguments can be used to
estimate the expected relation between cluster mass and temperature,
but the exact cosmological dependence could be sensitive to merger
rates as a function of cosmology or other non-linear effects.

Furthermore, the mapping between the initial density field and the
number density of clusters of a given mass as a function of redshift,
i.e., the mass function, is not known perfectly at this time. The
current generation of large cosmological simulations offers hope for a
much better understanding in coming years. An important use for a
large scale SZE cluster survey will be to characterize this mass
function and test the reliability of various analytical
\citep{press74,bond91,sheth01} and numerical \citep{jenkins01}
estimates.

In order to exploit fully the potential of SZE surveys, the limiting
mass as a function of redshift for galaxy clusters will have to be
understood to an accuracy of better than 5\% and uncertainties in the
mass function must be reduced to better than 10\%
\citep{holder01a}. The former will require a concerted observational
effort and is likely to be the most difficult to achieve; the latter
requirement is not far from current uncertainties
\citep{jenkins01}. Note that it is not required that we know the mass
of each cluster in the catalog to this accuracy, but only that we can
characterize the cluster detection efficiency as a function of mass to
this level.

While some of the theory of the physics of galaxy clusters is not
known, there are plenty of observational diagnostics which can be used
in the interpretation of SZE surveys. High resolution SZE imaging of
high redshift clusters will provide information on the relative
importance of gas dynamics \citep{holder01} to the observed properties
of galaxy clusters. The main effect of most gas processes should be to
cause the gas to be less centrally concentrated, either because the
low-entropy gas has been removed \citep{bryan00} or because the gas
has gained entropy from non-gravitational heating.

It may also be possible with the next generation of large format
multi-frequency bolometer arrays to exploit the small relativistic
corrections to the \sze\ spectrum to determine the gas temperature.
Such measurements will allow an understanding of temperature structure
in the gas, even without follow-up X-ray observations. A direct
SZE-weighted temperature would be more directly relevant to the SZE
observations in determining the gas mass, and comparisons with X-ray
spectral temperatures, when possible, could provide valuable
information on the temperature structure along the line of
sight. Temperature information, combined with high-resolution imaging,
allows a reconstruction of the cluster potential and therefore can
provide important constraints on the gas mass fraction as a function
of radius as well as a diagnostic for the effects of gas cooling.

In terms of the properties of the cluster catalog produced by a SZE
survey, it is difficult to avoid the conclusion that the survey limit
will be mainly dependent only on cluster mass and that this mass limit
will be relatively flat with redshift beyond $z \sim 0.1$. This will
make the resulting catalog especially useful for studies of the
physics of galaxy clusters. The requirements on understanding the
cluster mass selection function for such studies are much less strict
than those required for cosmological studies. The results from
detailed studies of individual clusters in the catalog will naturally
feed back and improve our understanding of the survey selection
function.

%%%%%%%%%%%%%%%%%%%%%%%%%%%%%%%%%%%%%%%%%%%%%%%%%%%%%%%%%%%%%%%%%%%%%%%%%%%%%
%%%  									  %%%
%%%	VII. Conclusions						  %%%
%%%  									  %%%
%%%%%%%%%%%%%%%%%%%%%%%%%%%%%%%%%%%%%%%%%%%%%%%%%%%%%%%%%%%%%%%%%%%%%%%%%%%%%

\section{SUMMARY}
\label{sec:summary}

The \sze\ is emerging as a powerful tool for cosmology. Over the last
several years, detection of the \sze\ toward massive galaxy clusters
has become routine,  as has high quality imaging at moderate
angular resolution of order an arcminute. Measurements of the effect already
have been used to place interesting constraints on the Hubble constant
and, through measurements of cluster gas mass fractions, the matter
density of the universe, $\Om$.

The next step is to exploit the redshift independence of the \sze\
signal to conduct blind surveys for galaxy clusters.  The limit for
such a survey is essentially a mass limit that is remarkably uniform
with redshift. The cluster catalog from such a unbiased survey could
be used to greatly increase the precision and redshift range of
present \sze\ constraints on the Hubble constant and $\Om$, and could,
for example, allow $\Da(z)$ to be determined to high redshift ($z \sim
2$).

The most powerful use of the SZE for cosmology will be the measurement
of the evolution of the abundance of galaxy clusters. SZE surveys are
ideally suited for this since they are able to probe the abundance at
high redshift as easily as the local universe. The evolution of the
abundance of galaxy clusters is a sensitive probe of cosmology. For
example, the yields from a deep \sze\ survey covering only ten square
degrees would be able to place interesting constraints on $\Om$,
$\Ol$, and $\sigma_8$.

A generic prediction of inflation is that the primordial density
fluctuations should be Gaussian.  Non-Gaussianity in the form of an
excess of high mass clusters should be readily apparent, especially at
high redshift, from \sze\ survey yields. SZE cluster surveys will
therefore probe both the structure formation history of the universe
and the nature of the primordial fluctuations. In this way, cluster
surveys are emerging as the next serious test of the cold dark matter
paradigm.

Current \sze\ observations, while routine, require substantial
integration time to secure a detection; a prohibitively long time
would be required to conduct blind surveys over a large region of sky
with the instruments now available. However, the next generation of
instruments now being built or planned will be substantially faster.
Dedicated interferometric arrays being built will be able to conduct
deep SZE surveys over tens of square degrees. Heterogeneous arrays,
such as the SZA combined with the OVRO array, will also allow detailed
high resolution follow up \sze\ observations of the resulting cluster
catalog.

A dedicated, low noise, single dish telescope with $\sim 1'$
resolution, equipped with a next generation, large format bolometric
array receiver ($\sim 1000$ elements) and operating from a superb site
would be able to conduct a deep \sze\ survey over thousands of square
degrees. The statistics provided by the yields from such a large
survey ($\sim 10^4$ clusters) in the absence of systematic effects and
assuming redshifts are known would be sufficient to determine precise
constraints on $\Om$, $\Ol$, $\sigma_8$, and even set meaningful
constraints on the equation of state of the dark energy.

The possible systematics that could affect the yields of \sze\
surveys are presently too large to realize the full potential of a
deep \sze\ survey covering thousands of square degrees. The
systematics include, for example, the uncertainties on the survey mass
detection limit due to unknown cluster structure and cluster gas
evolution, as well as the uncertainties in the theoretical mapping
between the initial density field and the number density of clusters
of a given mass as a function of redshift, i.e., the mass function.

These systematics can begin to be addressed through detailed follow-up
observations of a moderate area \sze\ survey (tens of square
degrees). High resolution \sze, X-ray, and weak lensing observations
will provide insights into evolution and structure of the cluster
gas. Numerical simulations directly compared and normalized to the
\sze\ yields should provide the necessary improvement in our
understanding of the mass function.

It is not unreasonable to consider the possibility of a space-based
telescope operating at centimeter through submillimeter wavelengths
with high angular resolution ($<1$ arcminute) and good spectral
coverage.  For studies of the SZE, this would allow simultaneous
determinations of electron column densities, temperatures, and
peculiar velocities of galaxy clusters. Such a satellite would make
detailed images of the cosmic microwave background, while also
providing important information on the high frequency behavior of
radio point sources and the low frequency behavior of dusty
extragalactic submillimeter sources.  The upcoming {\it Planck
Surveyor} satellite is a first step in this direction; it should
provide an SZE all-sky survey although at moderate, $\sim 5$
arcminute, resolution. Such a survey should find on the order of
$10^4-10^5$ clusters, most of them at redshift $z<1$.

We can look forward to the \sze\ emerging further as a unique and
powerful tool in cosmology over the next several years as the next
generation of \sze\ instruments come online and \sze\ surveys become a
reality.

\section*{ACKNOWLEDGEMENTS}

We thank 
M.\ Joy and 
W.\ Holzapfel for their considerable input to this review 
and
W.\ Hu,
S.\ LaRoque, 
A.\ Miller, 
J.\ Mohr, and 
D.\ Nagai for their comments on the manuscript. We also thank 
M.\ White and C.\ Pryke for assistance with Figure~\ref{fig:cmb-sze}.
This work was supported in part by NASA LTSA account NAG5-7986
and NSF account AST-0096913. JEC also acknowledges support from
the David and Lucile Packard Foundation and the McDonnell
Foundation. EDR acknowledges support from 
a NASA GSRP fellowship (NGT5-50173) and a Chandra Fellowship (PF1-20020). 

\begin{multicols}{2}

%\begin{thebibliography}{148}

\end{multicols}

\end{document}